\font\msamstex=msam10
\def\gtrsim {\,\mbox{{\msamstex \char 38}}\,}

\documentstyle[
aps,prb]{revtex}

\begin{document}
\draft

\title{Conformations of dendrimers in dilute solution}

\author{Edward G.~Timoshenko\thanks{Corresponding author. 
Web: http://darkstar.ucd.ie; 
E-mail: Edward.Timoshenko@ucd.ie}
}
\address{
Theory and Computation Group,
Centre for Synthesis and Chemical Biology,
Conway Institute for Biomolecular and Biomedical Research,
Department of Chemistry, University College Dublin,
Belfield, Dublin 4, Ireland}

\author{Yuri A.~Kuznetsov}
\address{
Centre for High Performance Computing Applications, 
University College Dublin,
Belfield, Dublin 4, Ireland}

\author{Ronan Connolly}
\address{
Theory and Computation Group,
Department of Chemistry, University College Dublin,
Belfield, Dublin 4, Ireland}

\date{\today}
\maketitle

\begin{abstract}
Conformations of isolated homo-- dendrimers of $G=1-7$ generations
with $D=1-6$ spacers have been
studied in the good and poor solvents, as well as across the
coil--to--globule transition, by means of
a version of the Gaussian self--consistent (GSC) method and
Monte Carlo (MC) simulation in continuous space 
based on the same coarse--grained model. The latter includes harmonic
springs between connected monomers and the pair--wise Lennard--Jones
potential with a hard core repulsion.
The scaling law for the dendrimer size, the degrees of bond stretching
and steric congestion, as well as the radial density, static
structure factor, and asphericity have been analysed.
It is also confirmed that while smaller dendrimers have a dense
core, larger ones develop a hollow domain at some separation
from the centre.
\end{abstract}

\vskip 1cm

\section{Introduction}
\label{sec:intro}

Dendrimers represent a class of well defined hyper--branched
macromolecules \cite{AdvPolHult} which can
be synthesised via a sequence of
carefully controlled repetitive reactions producing
regular structures \cite{DendrimerReview1}.
There are high aspirations \cite{Tomalia,Frechet} of chemists 
and physicists regarding applications of
such novel polymers as advanced materials
for medicinal  \cite{Bosman} uses, superior catalysts 
\cite{DendrimerReviewCatalysis,Groot}, 
and drug delivery vehicles \cite{Liu}, to mention just a few. 
Importantly for such applications, dendrimers are quite flexible molecules, 
which have a well accessible interior part as well as a large
exterior surface. Furthermore, dendrimers
can controllably change their size and density distribution 
under the influence of external conditions.
Some of the potential applications of dendrimers moreover 
require a hollow interior, which could be used for accommodating guest
molecules \cite{GuestMol}.
In addition, spacers, charges, and distinct types of monomer units 
should allow one to further fine tune particular behaviour of dendrimers.

Although the ideal model of dendrimers with the 
harmonic springs or random walks
can be solved analytically \cite{LaFerla}, 
such results are of a limited value since the excluded volume interactions
are of paramount importance for dendrimers \cite{AdvPolFreire}.
However, including these in order to describe the range of the 
coil--to--globule transition in sufficient detail is a considerable challenge
even for the homo-- dendrimer. This problem has 
attracted a number of researchers yielding a significant body of
theoretical and computational work in recent years,
with some experimental results 
\cite{ExperDendri} available to assist them. 


De Gennes and Hervet \cite{deGennesHervet}
have performed a seminal calculation using 
a version of the Edwards' self-consistent field method
upon an assumption that an onion--like layered distribution 
of the dendrimer generations applies.
They have predicted a density minimum at the core of the dendrimer
and the scaling law for the radius of gyration, 
${\cal R}_g \sim N^{1/5}$, where $N$ is the total number of monomers,
both of which henceforth have been debated.
Later on Boris and Rubinstein \cite{BorisRubinstein} have proposed a
more complicated self--consistent mean--field and related Flory--type theories.
They have found that the density is
monotonically decreasing with a maximum at the core. 
However, such a theory does not contain spatial correlations and 
thus may have considerable limitations in predicting the internal structure
of dendrimers.

A version of the Gaussian variational theory
\cite{Ganazzoli-CondMat}, which relies on a somewhat problematic
\cite{ConfTra} virial representation of the excluded volume
interactions \cite{Edwards,CloizeauxBook}, has been used
by Ganazzoli {\it et al} for
describing the size, shape and intrinsic viscosity of dendrimers
in a good solvent \cite{Ganazzoli-DendriTerra}. Based on this
approach the dynamic properties such as the relaxation times spectrum, 
dynamic structure factor and the viscoelastic moduli,
\cite{Ganazzoli-DendriDynamics},
as well as the inter--chain theta--temperature \cite{Ganazzoli-DendriUnper} 
have been also investigated. Unfortunately, constraints of the
numerical procedure employed limited such studies to
dendrimers with $D=1,2$ spacers only.

A number of simulation studies using either Molecular Dynamics (MD) or
Monte Carlo (MC), both on and off- lattice, whether atomistically
detailed or coarse--grained, have been also performed.
Naylor {\it et al} \cite{TomaliaMD} have discussed the MD simulation
of PAMAM (poly(amido-amine)) dendrimers. This study, albeit generally
interesting, could cover a timescale of hundreds of picoseconds only.
Later Lescanec and Muthukumar \cite{Lescanec} have proposed a kinetic 
growth model in a 3-d off-lattice MC. 
They have reported the highest density at the core and have also 
found that the terminal monomers can traverse the dendrimer.
Their approach, however, was in essence a non--equilibrium one 
as sufficient equilibration was not been allowed.
This may explain a  rather small reported value
of the swelling exponent $\nu\simeq 0.22$
in the proposed scaling law ${\cal R}_g \sim N^{\nu}\,D^{1/2}$
for the good solvent. 
Next, Mansfield and Klushin  \cite{MansfieldKlushinMacromol}
have performed a MC simulation on 
a diamond lattice with $D=7$ spacers modelled as self--avoiding walks. 
Their paper has managed to overcome some of the previous limitations and
it contained some important insights.
Nevertheless, the $D$ dependence has not been investigated there.
Furthermore, a possibility of lattice artefacts for the 
restricted branch points may raise some doubt in the more refined 
features of their predictions.
Chen and Cui \cite{ChenCui} have performed
off--lattice MC simulations for a good solvent in a 
model with freely rotating bonds. They have also attempted to fit the
radius of gyration as 
$R_g \sim D^{\nu}\,2^{(2\nu-1)G}\,G^{1-\nu}$, i.e. 
in terms of both $D$ and the 
number of generations $G$.
While the value of $\nu$ has been related to the 
inverse fractal dimension of the open/ring chain, just as in 
Ref.~\onlinecite{Biswas}
containing the renormalisation group calculation by Biswas and Cherayil,  
the meaning of the $G$-dependent part of the scaling law 
lacked such clarity.

Further MD simulation by Murat and Grest \cite{MuratGrest} has 
relied on a coarse--grained model with the Lennard--Jones non--bonded
and the FENE bonded interactions
employing a white noise for the solvent and temperature effects.
This simulation involved sufficient equilibration times and
has yielded a number of interesting results. 
For instance, the dendrimer size at a fixed number of spacers $D$ 
in a good solvent
was found to scale as $R_g \sim N^{\nu}$ with the
value of the exponent $\nu$ surprisingly close to $1/3$.
Interestingly, the indication about a hollow domain inside the dendrimer, 
which has been first noted in Ref.~\onlinecite{MansfieldKlushinMacromol}, 
has been reiterated and further elaborated by performing the partial 
densities analysis over generations.
These authors have also looked at the dendrimer contraction on changing the
solvent quality from good to poor. They have concluded that 
the law ${\cal R}_g \sim N^{1/3}$ (at fixed $D$)
applies for the more compact globular dendrimer, which seemed reasonable
from the point of view of the space filling argument. 

Despite this considerable progress, understanding of
the homo- dendrimers is not as yet fully satisfactory.
First, a reasonably large span of spacer and generation numbers under
variable solvent conditions has until now been hard to study. Second, 
it was difficult to achieve both a reliable equilibration and a
good averaging statistics afterwards at the level of
affordable computational times.
Clearly though,
these requirements must be met at the same time in order
to be able to extract reliable scaling predictions and to make
firm conclusions on the intricacies of the internal structure
of dendrimers. Moreover, not only the good, but the theta- and poor solvents,
as well as the thermodynamics of the coil--to--globule, are of interest.  
Perhaps, particular deficiencies of the techniques employed, an insufficient
accuracy, and a limited range of systems
may explain some of the discrepancies 
which remained between the results of different approaches.
Therefore, we believe that it would be important to revisit the problem.

Here we shall try to re--examine the fundamentals of dendrimers
in a broader context of their overall conformational
behaviour as a function of the solvent quality.
On the one hand, this may help to resolve some of the remaining 
controversy in the concrete predictions.
On the other hand, it may
shed some new light on the problem thanks to studying the
observables which hitherto have not been investigated and 
to performing a type of analysis which is more novel. 
Last but not least, one may hope to 
improve the statistics and to study a broader range of dendrimers
by using a well optimised MC code 
\cite{Torus,CopStar,CorFunc} on faster modern computers \cite{Computers}.

Our recent progress \cite{GSC0} on extending
the Gaussian self--consistent (GSC) method may 
offer some advantages and permit comparisons
to be made with the simulation data. 
The improved technique avoids the use of a virial expansion, 
but instead operates directly with a given Hamiltonian involving any pair--wise
bonded and non--bonded interactions, just as in direct simulations.
Ref. \onlinecite{GSC0} deals with all the technical details
and it contains a comprehensive comparison with the MC data
performed for open, ring and star homopolymers with variable flexibility 
in a good and poor solvents. 
The method appears to be relatively accurate
for a number of observables and its limitations are well understood.
Because the new GSC technique is computationally very
fast and since it directly yields the
equilibrium averages for the observables of interest, 
it will allow us to consider a larger range of system sizes 
and a broader region of solvent conditions than with the
equivalent MC simulation.  

\section{Model and notations}
\label{sec:model}

The current coarse--grained homo-- dendrimer model is based
on the following Hamiltonian (energy functional) 
\cite{CorFunc,CopStar,CombStar} 
in terms of the monomer coordinates, ${\bf X}_i$:
\begin{eqnarray}
\frac{H}{k_B T} & = & \frac{1}{2\ell^2}  \sum_{i\sim j} \kappa_{ij}
        ({\bf X}_i - {\bf X}_{j})^2
+\frac{1}{2} \sum_{ij,\ i\not= j}
U^{(lj)} (|{\bf X}_i - {\bf X}_j|).
\label{cmc:hamil}
\end{eqnarray}
Here the first term represents the connectivity of the dendrimer with
harmonic springs of strength $\kappa_{ij}$ introduced between
any pair of connected monomers (which is denoted by $i\sim j$).
The second term represents 
pair--wise non--bonded, specifically van der Waals interactions
between monomers.
We shall adopt the Lennard--Jones form of the potential,
\begin{equation}\label{VLJ}
U^{(lj)}(r) = \left\{
\begin{array}{ll}
+\infty, &  r < d^{(0)}\\
U^{(0)} \left( \left( \frac{d^{(0)}}{r}\right)^{12}
- \left( \frac{d^{(0)}}{r} \right)^{6} \right), &
r > d^{(0)}
\end{array}
\right.,
\end{equation}
with hard core part and the monomer diameter $d^{(0)}$,
where $U^{(0)}$ is the dimensionless strength of the interaction. 

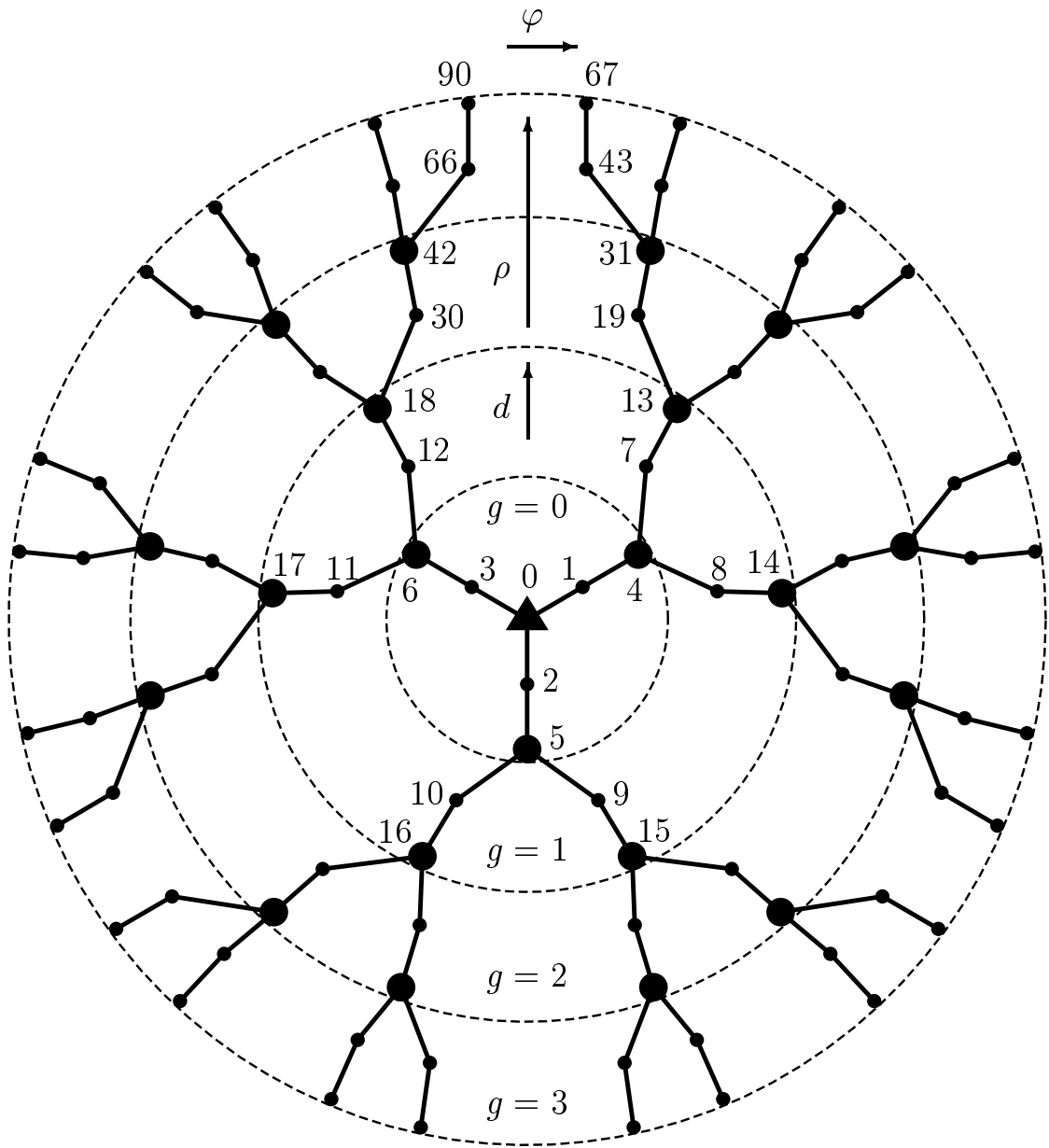
\begin{figure}
\caption{ \label{fig:dendri}
Schematic connectivity diagram of the dendrimer D2G3.
Here dashed circles denote different generations,
big filled circles correspond to the branch points
and the triangle is the core monomer.
Notations for the indices $d,\rho,\varphi$ and for some selected monomers,
which are numbered in the clock--wise and outwards manner, are
also introduced.
}
\end{figure}

It is important to introduce convenient notations.
Let us consider a dendrimer of $G$ generations, each of which consists
of $D$ spacers, with $F_0$ functional core 
(corresponding to the generation $g=-1$) and $F$ functional other 
branch points (see Fig.~\ref{fig:dendri}).
Thus, if $D=a$ and $G=b$ we shall briefly denote such a
dendrimer as DaGb as in
Refs.~\onlinecite{Ganazzoli-DendriTerra,Ganazzoli-DendriDynamics}.
Then the total number of monomers will be, 
\begin{equation}
\label{N}
N=1+F_0 D \frac{(F-1)^{G+1}-1}{F-2}.
\end{equation}
Note that below we shall consider results for a particular example 
of tri--functional branching $F=F_0=3$. 
It is convenient to represent monomer indices via triads
$i=(g,d,\varphi_g)$, where $g=0,1,\ldots G$ is the generation index,
$d=0,1,\ldots D-1$ is the spacer index, and $\varphi_g = 0, 1,
\ldots F_0 (F-1)^g -1$ is the angular in--generation index. 
It is also useful to define the shell radial index via
$\rho = D g + d +1$, and hence instead of the triad, one can use
the `polar coordinates' index notation $i=(\rho,\varphi_g)$.
Both of these will be used interchangeably henceforth.

\section{Techniques}
\label{sec:techniques}

\subsection{Monte Carlo technique}

We use the Monte Carlo (MC) technique with the standard Metropolis 
algorithm \cite{AllenTild} and local monomer moves, 
based upon the implementation 
described by us in Refs.~\onlinecite{Torus,CorFunc}.
The position of a randomly picked 
monomer in the spherical coordinates is displaced by
$\Delta r = \delta\,r_1$, $\Delta\theta=\pi\,\theta_1,$
$\Delta\phi=2\pi\,\phi_1$
where $(r_1,\theta_1,\phi_1)$ is a triple of independent 
standard uniform deviates and $\delta$ is an additional
parameter of the MC scheme
characterising the timescale involved in the Monte Carlo sweep (MCS),
the latter being defined as $N$ attempted MC steps.
The Metropolis check, $\Delta E \leq -k_B T \ln {\sf r}_{1}$,
where ${\sf r}_{1}$ is yet another standard
uniform deviate, is then used as a criterion for accepting 
trial new conformations. As a result, the ensemble of independent
initial conformations would
converge to the Gibbs equilibrium distribution with increasing
number of sweeps.
Initial conformations of dendrimers were taken as planar 
and rather overstretched ones (akin to the schematic
representation in Fig.~\ref{fig:dendri}),
but then they were subjected to extensive equilibration for a required time
before any simulation was commenced.

To ensure good equilibration, the
behaviour of global observables such as
the energy and radius of gyration was monitored. 
Upon reaching equilibrium  these cease to have 
a global drift and start exhibiting characteristic
fluctuating behaviour around well defined mean values.
Then, depending on the dendrimer size,
about $Q=2\cdot 10^4$ of statistical measurements 
have been performed, typically separated by about $40\, N^2$ of attempted 
Monte Carlo steps to ensure statistical independence of sampling. 
The mean value and error of sampling of an observable $A$ are then
given by the mean 
$\langle A \rangle=(1/Q)\sum_{\gamma}^Q A_{\gamma}$
and by $\pm \sqrt{(\Delta A)^2/Q}$ respectively.

\subsection{The Gaussian Self--Consistent method}

The main objects in the GSC method are the mean--squared distances
between monomers,
\begin{equation} \label{eq:Ddef}
{\cal D}_{ij}(t) \equiv \frac{1}{3}\biggl\langle
({\bf X}_{i}(t) - {\bf X}_{j}(t))^2 \biggr\rangle.
\end{equation}
The GSC method is based on replacing the stochastic ensemble for
$\bbox{X}_i$ with the exact Hamiltonian in 
the Langevin equation of motion onto the trial ensemble 
$\bbox{X}_i^{(0)}(t)$ with a trial
Hamiltonian $H^{(0)}(t)$, which is a generic quadratic form with
matrix coefficients called the time--dependent effective potentials,
\begin{equation} \label{gsc:trial}
H^{(0)} [{\bf X} (t)] = \frac{1}{2}\sum_{ij} V_{ij} (t)\,
{\bf X}_i (t)\, {\bf X}_{j} (t).
\end{equation}
Then one requires that the inter--monomer correlations satisfy the condition,
\begin{equation}
\langle \bbox{X}_{i}(t) \bbox{X}_{j}(t) \rangle_0 =
\langle \bbox{X}_{i}^{(0)}(t) \bbox{X}_{j}^{(0)}(t) \rangle_0,
\end{equation}
which means that the trial ensemble well approximates the exact one.
This yields 
the self--consistent equations \cite{CopStar},
which in the absence of the hydrodynamic interactions are,
\begin{equation} \label{eq:kinmain}
\frac{\zeta_b}{2}\frac{d}{dt}{\cal D}_{ij}(t) = -\frac{2}{3}
\sum_{k}({\cal D}_{ik}(t)-{\cal D}_{jk}(t))\left(
\frac{\partial {\cal A}[{\cal D}(t)]}{\partial {\cal D}_{ik}(t)}-
\frac{\partial {\cal A}[{\cal D}(t)]}{\partial {\cal D}_{jk}(t)}
\right).
\end{equation}
Here $\zeta_b$ is the friction coefficient of a monomer
and the instantaneous free energy has the same functional
expression via the instantaneous ${\cal D}_{ij}(t)$ as it
has at equilibrium.

The expression for the free energy,  
${\cal A} \equiv  {\cal E}^{(tot)} - T {\cal S}$
has been given in Ref. \onlinecite{GSC0}.
We would not like to reproduce that expression here as it is 
somewhat cumbersome.
However, it should be mentioned that
the total mean energy includes the energy of the bonded and
non--bonded interactions ${\cal E}$ as well as the hard--core
contribution as follows,
${\cal E}^{(tot)}= {\cal E} + {\cal A}^{(hs)}$.
The second term is given by the generalised Carnahan--Starling term
\cite{CarnahanStarling,GSC0},
\begin{eqnarray}
\frac{{\cal A}^{(hs)}}{k_B T}  & = & \sum_i 
     \frac{\eta_i (4 - 3\eta_i)}{(1 - \eta_i)^2}, 
\qquad
\eta_i  =  \sum_{j \neq i} F^{(\eta)} 
\left[ \frac{\sqrt{{\cal D}_{ij}}}{d^{(0)}} \right],
\label{eq:etai} 
\\
F^{(\eta)} [y]  & = & \frac{y\,{\rm erf}\left(\frac{1}{y\sqrt{2}}\right) -
                \sqrt{2/\pi}\,\exp\left(-\frac{1}{2y^2}\right)}{8y},
\label{eq:Fy}
\end{eqnarray}
where $\eta_i$ is known as the packing coefficient of the monomer $i$,
which would have an important role below.

The stationary limit of the GSC Eqs. (\ref{eq:kinmain})
produces the equations for
the minimum of the free energy, which are the same as
those derived from the Gibbs--Bogoliubov
variational principle.
Although here we shall only be concerned with the equilibrium,
the numerical solution of Eq. (\ref{eq:kinmain}),
applied until the stationary limit is reached,
presents one of the most efficient techniques
for finding the global free energy minimum.
This, based on the fifth order adaptive step Runge--Kutta
integrator \cite{ConfTra}, is used for obtaining the results
from the GSC technique.

It is also important to map out all equivalent pairs of the matrix
${\cal D}_{ij}$, which is discussed in Appendix \ref{app:A}
based on the concept of the topological tree,
so that only $C_{ind}^{(2)}=(G+1)(D/2)(D(G^2/3+(13/6)G+2)+2+G/2)$
independent elements out of total $C_{tot}=N(N-1)/2$ remain in it
due to the kinematic symmetries of the dendrimer.
It is interesting to remark, more generally, 
that exactly the same symmetries will
exist for a hetero-- dendrimer as long as the monomers at
each shell index $\rho$ are identical to each other. 

The kinematic symmetries for the matrix ${\cal D}_{ij}$
also yield analogous symmetries for the matrix
$V_{ij}$.
Thus, the computational expenses
per step in our calculations are of order $t_c \sim N\,C^{(2)}_{ind}$, 
where $C^{(2)}_{ind} \sim D^2 G^3/6$
is the total number of independent elements in the ${\cal D}_{ij}$ matrix
(see Eq.~(\ref{degInd})).
These symmetries significantly reduce the computational times
compared to the equivalent system in the MC method,
where such symmetries only appear in the observables
after averaging over the statistical ensemble. 
For comparison, the computational expenses per step in 
MC are of order $t_c \sim N\,\Delta t\,S$,
where $\Delta t\sim N^2$ is the number of MC
steps needed to ensure good statistical independence
between measurements, and $S$ is the number of measurements
needed for sampling of observables. Typical values of $S$
should be of order of $10^4-10^5$ for good accuracy
\cite{CorFunc}.
For example, for D3G5 dendrimer (consisting of $N=568$ units) 
in the good solvent, $U^{(0)}=1$,
the GSC method takes about $22$ minutes \cite{Computers} to
reach equilibrium and produce its data, whereas the MC 
manages to make only $100$ measurements of observables per hour 
after required equilibration.
Thus, the GSC is about $10^3$ times faster than the MC for
producing the same results here.

\section{Results}\label{sec:results}

We shall restrict ourselves to springs of equal strength,
$\kappa_{ij} = 1$, and choose the hard sphere diameter 
equal to the statistical length, $d^{(0)}=\ell$, in Eq. (\ref{cmc:hamil})
as in Ref. \onlinecite{CorFunc}.
Moreover, henceforth we shall use the mean energy ${\cal E}$
expressed in units of $k_B T$ and the mean--squared
distances ${\cal D}_{ij}$ and mean--squared 
radius of gyration, $3 {\cal R}_g^2$, expressed
in units of $\ell^2$.

In our figures below we shall
present data from MC simulation via thick
lines and empty circles, whereas data from GSC theory via thin
lines and small filled circles.
In this work we have analysed the following dendrimer sizes:
$G=1-7$ for $D=1-3$ and $G=1-6$ for $D=4-6$.

\subsection{Good solvent}\label{sec:good}

We shall start by examining the dendrimer overall
size and structure in the good athermal solvent, $U^{(0)}=0$.

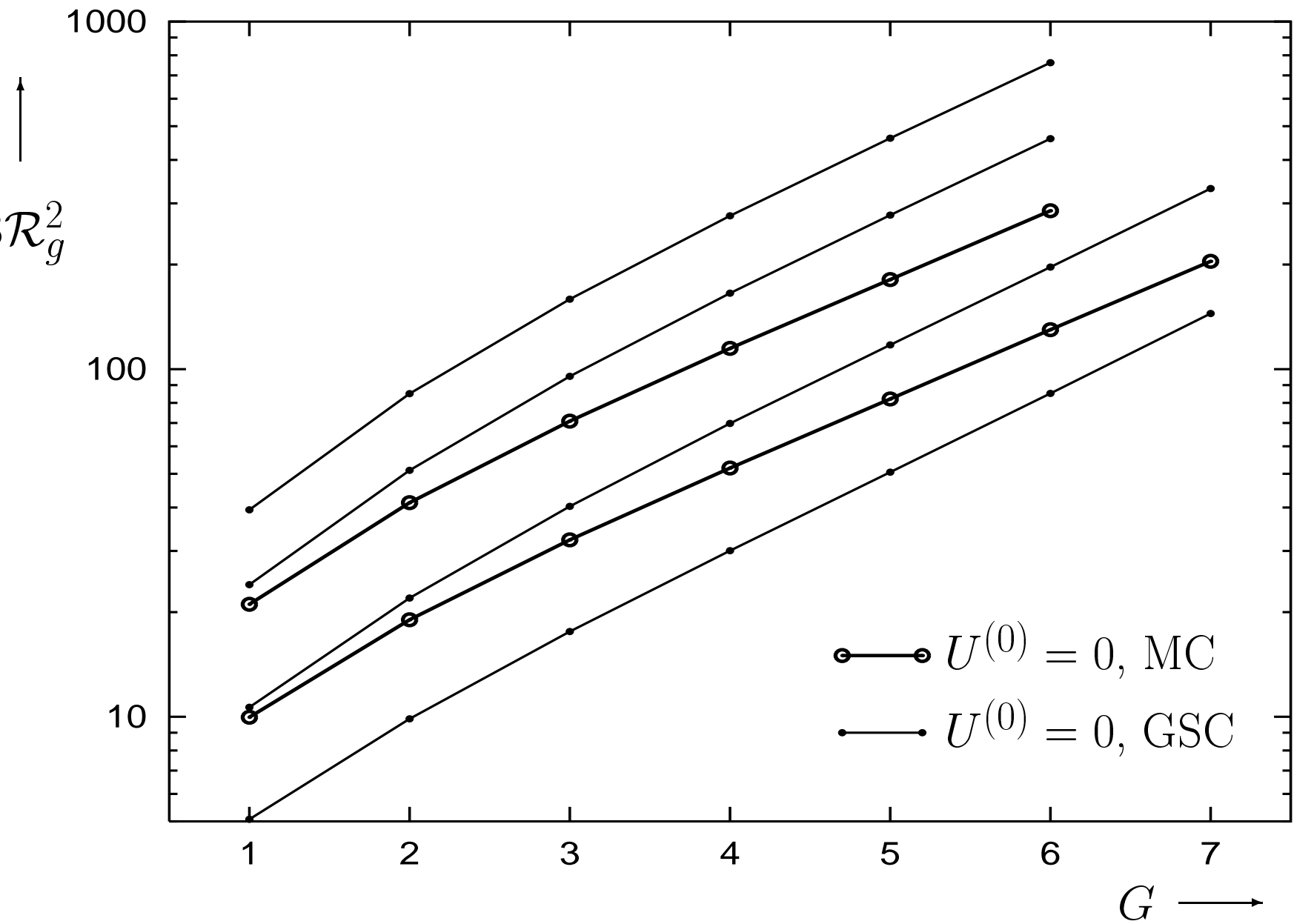
\begin{figure}
\caption{ \label{fig:RgCoilG}
Plots of the mean--squared radius of gyration
$3{\cal R}_g^2$ vs the number of generations $G$
for the dendrimers in the good solvent ($U^{(0)}=0$)
from the GSC theory (filled circles and thin lines) and
from the MC simulation (empty circles and thick lines).
The curves (from bottom to top) correspond to the following values:
$D=1$ (GSC), $2$ (MC), $2$ (GSC), $4$ (MC), $4$ (GSC), $6$ (GSC).
The largest relative statistical error, 
$\delta R_g^2=\Delta R_g^2/R_g^2 \cdot 100\%$,
in this plot occurs for the D2G7 dendrimer and equals $0.1\%$.
}
\end{figure}

The mean--squared radii of gyration $3{\cal R}_g^2$
of dendrimers with varying number of spacers $D$ vs
the number of generations $G$ are depicted in Fig.~\ref{fig:RgCoilG}.
Clearly, $3{\cal R}_g^2$ increases with both $D$ and $G$, but in
a rather different manner, which will be discussed in full detail 
below in Subsec.~\ref{sec:scaling}. Corresponding MC and GSC curves 
follow each other quite closely, but the GSC theory increasingly
overestimates the dendrimer size with increasing $D$ and $G$,
in a manner similar to our discussion in Ref.~\onlinecite{GSC0}.

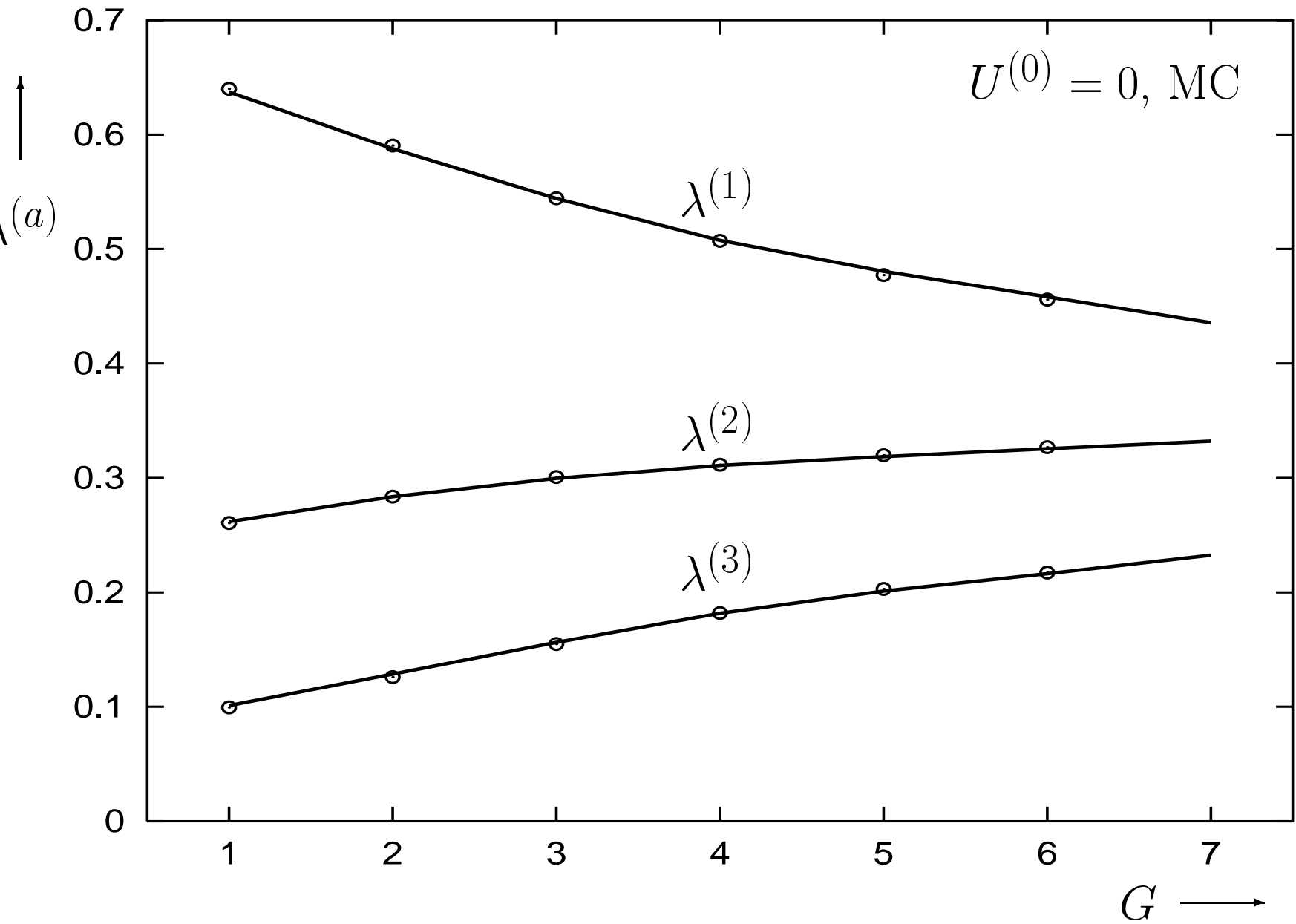
\begin{figure}
\caption{ \label{fig:lambda}
Plots of the mean ratios of the eigenvalues of the shape tensor 
$\lambda^{(a)}$ vs the number of generations $G$ for the dendrimers 
in the good solvent ($U^{(0)}=0$) from MC simulation.
Solid curves correspond to $D=2$ and empty circles to $D=4$.
The largest relative statistical errors, 
$\delta \lambda^{(a)}=\Delta \lambda^{(a)}/\lambda^{(a)} \cdot 100\%$,
in this plot occur for the D2G7 dendrimer and are equal to:
$\delta\lambda^{(1)} = 0.15\%$,
$\delta\lambda^{(2)} = 0.2\%$ and $\delta\lambda^{(3)} = 0.3\%$.
}
\end{figure}

As $G$ increases, dendrimers become more spherical, which is
illustrated via the average normalised `axes of inertia' $\lambda^{(a)}$
defined by Eqs.~(9,10) in Ref.~\onlinecite{CorFunc}. These converge
towards a perfect sphere limit, $\lambda^{(a)}=1/3$, as $G$ grows
in Fig.~\ref{fig:lambda}.
Interestingly, these quantities seem to be practically independent
of the number of spacers $D$, which is illustrated by using
circles ($D=4$) and lines ($D=2$) in the figure.

Typical snapshots of dendrimers D3G3 and D3G7 in Figs.~\ref{fig:Snap}
from MC simulation do confirm the observation that dendrimers
become more spherical as $G$ increases. Generally, a fairly large
dendrimer tends to fill in the available space quite effectively and 
relatively uniformly, although some local dilutions and accumulations remain.
The size of the dendrimer is well defined with very few branches 
outstretching beyond it.

The shape of the static structure factor (SSF) in the Kratky representation
in Fig.~\ref{fig:Kratky} for small $G$ is
reminiscent of that of a ring, or a star with few
arms (see Figs.~9,10 in Ref.~\onlinecite{CorFunc}), but
tends more to that of a globule with increasing number of generations.
Even though the tail of the plot still increases with $\hat{p}$,
it develops a characteristic oscillating behaviour. This is consistent
with the trends in $\lambda^{(a)}$ above, particularly so since 
the $D$-dependence is almost unnoticeable in $\hat{S}$ as well.

The total and partial densities, 
\begin{equation}
\label{DensDef}
g^{(1)}_g(r)\equiv 
\sum_{i\in g}\langle \delta({\bf r}_i-{\bf r}_{cm}-{\bf r})\rangle, 
\quad 4\pi\int_0^{\infty} r^2\,dr\,g^{(1)}_g(r)=N_g=F_0 D (F-1)^g.
\end{equation}
of the dendrimers in Figs.~\ref{fig:DensD2G} allow one to understand
the monomer distributions in more detail. In a relatively small-$G$ 
dendrimer (Fig.~a) the total density monotonically decreases with the 
separation from the centre--of--mass ${\bf r}_{cm}$. The highest density
at the origin also occurs for the partial density of inner generations,
and it follows a similar pattern.  Remarkably, the terminal generation
density is rather delocalised spreading evenly from the outer boundary towards
the very centre of the dendrimer. This can only be rationalised by allowing
a possibility of the ends re-entering the dendrimer 
and reaching far inside, although the density of the terminal generation
exceeds that of the others somewhat near the very edge of the dendrimer.

These features are further complicated for dendrimers of a greater
number of generations, typically for $G > 5$, at least in the considered
cases of $D=2,4$. While the behaviour of the terminal generation density
remains very similar in Fig~\ref{fig:DensD2G}b, the density of 
non--terminal generations, and hence the total density, develop a 
well noticeable dip at certain separation. This indicates that 
larger-$G$ dendrimers have the lowest density domain fairly
close to the centre, but not at the centre itself, which always has 
the highest density of all. We may note here also that while the
core monomer coordinate does differ from that of the centre--of--mass
somewhat, they are quite close to each other in a large-$G$ dendrimer.
Thus, this most hollow domain is not located where the core monomer is
and its separation from the core steadily increases with $G$.
Clearly, from Fig~\ref{fig:DensD2G}b one can see that the increase
of the total density at the origin comes entirely from the density
of the innermost generations.
However, the core domain is at its most dense in a small-$G$ dendrimer
and it becomes less dense for larger dendrimers (in terms of both
$G$ and $D$ actually) as can be seen by comparing the values
of $g^{(1)}(r=0)$ in
Figs.~\ref{fig:DensD2G}a and b. This `entropic pull' effect
is due to an increased entropy thanks to extra branches 
attached to a larger dendrimer, 
leading to a higher stretching of the springs near the core with larger $G$.

Indeed, in 
Figs.~\ref{fig:DrrCoilD2},\ref{fig:DrrCoilD4},\ref{fig:DrrCoilD6}
we depict the mean--squared distances between nearest neighbours along a
branch vs the shell index $\rho$. This is perhaps
the best way to characterise such stretching.
While the main bodies of Figs.~\ref{fig:DrrCoilD2},\ref{fig:DrrCoilD4}
present the data from MC simulation, the insets exhibit the same
results obtained from the GSC theory.
Clearly, the concrete numbers for 
${\cal D}_{(\rho-1,0)\,(\rho,0)}$ are somewhat overestimated by the GSC
theory compared to the MC data \cite{GSC0}, 
especially for small $\rho$ in
relatively large-$G$ dendrimers. Nevertheless, 
the particular patterns and overall shapes of the MC and GSC plots
are rather similar.

As can be seen from Fig.~\ref{fig:DrrCoilD2}, 
for small dendrimers D2G1-3,
the spring from the core to the first monomer is stretched most, 
from the first to the second one is less, and so on, with the function
${\cal D}_{(\rho-1,0)\,(\rho,0)}$ decreasing monotonically.
However, for dendrimers with $G>3$ the function develops a characteristic
descending step--like behaviour: the spring stretching within the
same generation is about the same, but it has a dramatic drop when passing
the branch point to the next generation. In all cases though, the bonds coming
from the core are stretched most, with this feature becoming stronger 
for larger $G$, whereas the terminal bonds are the least stretched of all,
with their length being nearly $G$-independent. 
The step--like behaviour can be easily interpreted by remarking that
the number of springs doubles at any branch point. Hence
the total unfavourable energy change due to a given magnitude
of bonds stretching would be more significant near the edges
than closer to the core. Note also that the
step--like dependence vanishes for the terminal generation.

Plots in Fig.~\ref{fig:DrrCoilD4} for $D=4$ spacers 
show that the stretching near the core increases significantly
with $G$ here as well.  One can also see that
the intra--generational steps develop a non--monotonic behaviour.
Bonds within a non--terminal generation first become shorter with 
increasing index $d$ until about $D/2$, then
grow longer again when approaching the next branch point. 
This U-pattern occurs because there is more steric congestion
near the branch points than in the middle of generations.
The terminal generation is the least stretched here as for $D=2$
and it has a similar monotonically diving behaviour ending up
at a $G$ independent stretching value. 
Similar plots for dendrimers with $D=6$ spacers in Fig.~\ref{fig:DrrCoilD6}
obtained from the GSC theory show the 
U-pattern becomes more pronounced as $D$ increases.

Finally, to address the issue of steric congestion we shall
analyse the behaviour of the monomer packing coefficient
$\eta_i$ defined by Eqs.~(\ref{eq:etai},\ref{eq:Fy}).
This quantity shows the volume fraction occupied by 
all other monomers around a given monomer number $i$, with the limit
$\eta_i \to 1$ corresponding to the fully packed situation
according to Eq.~(\ref{eq:etai}).

In Figs.~\ref{fig:EtaCoilG6},\ref{fig:EtaCoilD3} the packing
coefficients vs the shell index $\rho$ along a branch are
shown for different dendrimers with $G=6$ and $D=3$ respectively.
For the former case it is convenient to normalise the 
horizontal axis as $\rho/D$ so that branch points occur at
integer values $0,\ldots, G$ with $G+1$ corresponding to the terminus. 
In Fig.~\ref{fig:EtaCoilG6} different curves correspond to
different numbers of spacers $D$. The curve for $D=1$ gives a general
outline of $\eta_{\rho}$ behaviour. The packing coefficient, and
hence congestion, first increases, reaches a maximum and then
rapidly decreases towards the termini, which are less
crowded than the core domain. A larger number of spacers $D$ leads
to an U-pattern behaviour for $g < G$ and a rapid dive for the
terminal generation, similar to our discussion of Fig.~\ref{fig:DrrCoilD6}.
Again, the middle of generations, where $\eta_{\rho}$ has smooth
minima, has a lower steric congestion since 
the monomers have only two nearest neighbours along the chain there
compared to three for branch points, where $\eta_{\rho}$ has
sharp maxima. As for the terminus, it has a single nearest 
chain neighbour only, resulting in a pronounced minimum of
$\eta_{\rho}$. 

In Fig.~\ref{fig:EtaCoilD3} we likewise present the packing
coefficients for dendrimers with increasing number
of generations $G$ and the fixed number of spacers $D=3$.
The general behaviour of $\eta_{\rho}$ is similar here. 
For small $G<5$ the function $\eta_{\rho}$ oscillates around
a constant, dropping down dramatically on the terminal generation.
However, in larger dendrimers, $G\ge 5$, the function $\eta_{\rho}$
starts from much lower values for small $\rho$ and then increases
overall up to penultimate generations, dropping down for the
terminal $g$. Such behaviour is consistent with a lower
density domain near the core seen via the spatial density
distribution in Fig.~\ref{fig:DensD2G}b. 
For the D3G7 dendrimer, the lowest packing coefficient value occurs
at around the middle of the $g=0$ generation (monomers 1,2,3
in diagram Fig.~\ref{fig:dendri}, which are buried 
deep inside the dendrimer as in Fig.~\ref{fig:Snap}b).
Note also that here we have $\eta_{D(G+1)} > \eta_0$, indicating
the onset of overcrowding in the terminal generation, which
would become very severe for larger $G$.
Overall, the packing coefficient typically
changes nearly twice in its value for relatively large $D$
and $G=6$, whereas the range of change in $\eta_{\rho}$ increases 
further with increasing $G$.

Limited simulations with a smaller monomer diameter $d^{(0)}=0.3\,\ell$
have been also performed and similar results have been obtained. 
Dendrimers with such reduced repulsions are
more compact and, clearly, as $d^{(0)}$ is decreased further they change in 
a cross--over manner towards the behaviour of the ideal dendrimers.

\subsection{Coil--to--globule transition and poor solvent}\label{sec:poor}

Now we shall bring our attention to the coil--to--globule
transition. Upon increasing Lennard--Jones attraction $U^{(0)}$
the dendrimer contracts and the mean--squared radius of gyration
decreases as shown in Fig.~\ref{fig:RgColTrans}a.
The magnitude of this change is quite significant, being
over $20$ times for D6G5 dendrimer. As our plot is drawn in a single
logarithmic scale one can see that the region of maximal relative change 
in value of $3{\cal R}_g^2$ occurs at around $U^{(0)}\simeq 2.5$,
with the change being more pronounced in larger dendrimers.
Clearly, from this plot and from Fig.~\ref{fig:RgColTrans}b,
which depicts the derivative of the specific mean energy, one
can conclude that the transition is continuous (second--order like).

The quantity $d({\cal E}/N)/dU^{(0)}$ changes from one
asymptotic behaviour at small $U^{(0)}$ to another at large
arguments, reaching eventually a constant value corresponding
to the maximal packing. The maximal change in this quantity occurs
at $U^{(0)}_{{\cal E}'}\simeq 2.5$, which may be viewed as a reasonable
practical definition of the transition point \cite{GSC0}.
As the transition is continuous it has a finite width and hence
any such definition is somewhat arbitrary. However, as $N$ increases
the transition narrows and becomes sharper.
It would be easy to consider the
point of  vanishing of the non--bonded part of the 
second virial coefficient (binary integral),
$u^{(2)}_{LJ} \equiv (1/2)\int d{\bf r}(1-\exp(-U^{(lj)}(|{\bf r}|)))$, 
which occurs at about $U^{(0)}_{LJ}\simeq 1.386$.
Unfortunately, such a definition of the theta--point is unacceptable
as it produces a result which does not depend on
the polymer connectivity or size.
Instead, one has to consider the total $u^{(2)}$ with the
full account of the bonded and non--bonded interactions.
This, however, is technically quite complicated 
as the Mayer--Ursell expansion is not easy to apply to dendrimers
with both types of interactions involved. 
Thus, we shall use $U^{(0)}_{{\cal E}'}$
as a convenient numerical definition of a point which 
becomes close to the exact theta--point for sufficiently large $N$.

In Tab.~\ref{tab:3} we report the values of $U^{(0)}_{{\cal E}'}$ for
dendrimers with varied $D$ and $G$. Clearly, either increased number of
generations or spacers, and hence increased $N$, would lead to a lower 
value of $U^{(0)}_{{\cal E}'}$
(larger temperature $T_{{\cal E}'}$) of the transition.
Indeed, the point $U^{(0)}_{{\cal E}'}$ corresponds to the
compensation of the positive conformational entropy ${\cal S}^{(gau)}$
with the negative two--body ${\cal E}^{(lj)}$ energy growing as
$N^2$ and much faster than ${\cal S}^{(gau)}$ in $N$,
so that smaller $U^{(0)}_{{\cal E}'}$ would suffice for the compensation.
More non--trivially though,
if we compare, for example, D3G6 and D6G5 dendrimers, both having about the
same $N$, the larger $D$ system would have a lower $U^{(0)}_{{\cal E}'}$
(larger temperature $T_{{\cal E}'}$).

For linear or ring polymers one can alternatively 
define the theta--point of the
current model by finding a value of $U^{(0)}$ at which the
radius of gyration of the system matches that of the corresponding
ideal system ($d^{(0)}=0$). In Tab.~\ref{tab:3} such values
$U^{(0)}_{{\cal R}}$ are presented in parentheses. Clearly though,
this definition is inappropriate for dendrimers as $U^{(0)}_{{\cal R}}$
shifts deep into the globular area for dendrimers with sufficiently
large $G$.

The region beyond $U^{(0)}_{{\cal E}'}$ corresponds to a globular
conformation of the dendrimers, to discussion of which we shall
now turn. In Fig.~\ref{fig:RgGloblG} we display the mean--squared
radii of gyration for dendrimers with different number of spacers
$D$ vs the number of generations $G$. The solid lines correspond
to the globule with $U^{(0)}=7$, that is, well beyond the transition
region, whereas the dashed lines correspond to similar 
ideal dendrimers ($d^{(0)}=0$ and $U^{(0)}=0$). These grow
with $D$ at a fixed $G$, and vice versa in both cases.
Interestingly, while the globule is more compact
than the corresponding ideal dendrimer at relatively small $G$, for
D1G$>$6 this trend reverses. This is due to a significant role of
the repulsive part of the non--bonded potential in dendrimers
with excluded volume interactions and it shows a limited relevance
of the model of ideal dendrimers when applied to the coil--to--globule
transition. In fact, the scaling behaviour at the transition point
differs from that of the ideal dendrimer unlike in the simpler cases
of open or ring polymers. Such issues will be examined by us further
in Subsec.~\ref{sec:scaling}.

Next, we shall compare the mean--squared distances between nearest 
neighbours along a branch vs the shell index $\rho$
for the globule in Fig.~\ref{fig:DrrGlobD6} to that in the good solvent
in Fig.~\ref{fig:DrrCoilD6}.
The function ${\cal D}_{(\rho-1,0)\,(\rho,0)}$ here has a similar
U-pattern inside each generation, a dive at the terminal generation,
and it increases as $G$ is increased, just as before.
However, the descending stairway behaviour seen for the good solvent
is no longer present now even for large $G$, and the function
oscillates around a constant value. Thus, the
bond stretching near the core is reduced, and apart from the
intra--generational variation and the special role of the termini,
the globular dendrimers are fairly evenly stretched as well as have
uniform density (see inset in Fig.~\ref{fig:DensD2G}a).
Note also that the maxima occur at pairs of
subsequent chain indices $\rho$. This means that the mean--squared
distances from the pre--branch monomer to the branch point are
almost the same as from the branch point to the post--branch one,
which indicates similarity in overall bond stretching 
with increasing $\rho$ and hence an approximate
equivalence of all branch points in the globule. Therefore,
the connectivity structure of the dendrimer is less
significant and more locally manifested in the globular state.

Similar conclusions can be made also from the behaviour
of the monomer packing coefficient for globular dendrimers 
in Fig.~\ref{fig:EtaGlobD6}.
Again, $\eta_{\rho}$ only depends periodically on the
spacer index $d$, being nearly identical for all generations
but the terminal one. The peaks occur at the branch points
and the termini have the lowest packing coefficient of all
just as for the good solvent in Fig.~\ref{fig:EtaCoilD3}.
Note, however, that the actual magnitudes of $\eta_{\rho}$
are much larger for the globule than for the good solvent
and approach $1/2$.
Moreover, the mean value of $\eta_{\rho}$ in the globule
steadily increases with $N$ and there is no hollow domain in
the globular conformation even for $G>5$.
Indeed, the total density $g^{(1)}(r)$ of a globule
dendrimer shown in the insets of Figs.~\ref{fig:DensD2G}a,b
is practically uniform and rather
localised within the size of the globule, thus having
higher values than for the good solvent. Remarkably, the terminal
generations still can traverse the dendrimer even in such a dense
state.

\subsection{Scaling relations}\label{sec:scaling}

First, one can fit the mean--squared radii of gyration at a fixed number of
generations $G$ in terms of $N$ expressed via $D$ by Eq.~(\ref{N}) as,
\begin{eqnarray}
\mbox{Good solvent:}\qquad\sqrt{3 {\cal R}_g^2}  & = & a_G\, N^{\nu_G},
   \label{eq:fitFlory}\\
\mbox{Theta--solvent:}\qquad\sqrt{3 {\cal R}_g^2}  & = & {\sf a}_G\, N^{{\sf v}_G},
   \label{eq:fitTheta}\\
\mbox{Poor solvent:}\qquad\sqrt{3 {\cal R}_g^2}  & = & A_G N^{\upsilon_G}.
\label{eq:fitGlobule}
\end{eqnarray}
The values of the exponent $\nu_G$ obtained from fitting of the data
are presented in Tab.~\ref{tab:1},
\begin{eqnarray}
\label{NuGF}
&& \mbox{MC:}\  \nu_G \simeq 0.58 \pm 0.02, \\ 
&& \mbox{GSC:}\ \nu_G \simeq 0.62 \pm 0.01.
\end{eqnarray}
Note that $\nu_G$ here practically coincides with the inverse
fractal dimension of an open repulsive chain.
Thus, we can interpret $\nu_G$ as
the Flory swelling exponent,
the value of which for large $N$ is more accurately 
known from the renormalisation group
calculation, namely $\nu_F=0.588 \pm 0.001$.
As discussed in Ref.~\onlinecite{GSC0} the GSC theory overestimates
the Flory swelling exponent by about $0.04$ as compared to the MC data
for this range of finite polymer sizes. This is consistent with the
amount by which the GSC values for $\nu_G$ exceed those of the MC
simulation. 

At the same time, the prefactor $a_G$ does not obey a power law in $G$,
but it can be fitted well via the following expression,
\begin{equation}
\label{aD}
\mbox{GSC: } \ a_G=(0.74\pm 0.01)\,2^{-\frac{G+1}{4.07\pm 0.1}}, \qquad
\mbox{MC: } \ a_G=(0.9\pm 0.03)\,2^{-\frac{G+1}{3.9\pm 0.2}}.
\end{equation}
Or by using Eq.~(\ref{N}) again,
we can express this for the good solvent via, 
\begin{eqnarray}
\sqrt{3 {\cal R}_g^2} & \sim & \left(\frac{3D}{N}\right)^{1/4}N^{\nu_F}
\sim D^{1/4}\, N^{\nu_F-1/4}, \\
\sqrt{3 {\cal R}_g^2} & \sim & D^{\nu_G}\,(F-1)^{\nu_D\,(G+1)},
\qquad \nu_D \approx \nu_G-\frac{1}{4}.
\label{RgNu}
\end{eqnarray}
Indeed, as can be seen from Tab. \ref{tab:2}, the resulting exponent
$\nu_D \simeq 0.37 \pm 0.01$ in the GSC method, whereas from MC the
resulting exponent value is $\nu_D \simeq 0.335 \pm 0.008 \gtrsim 1/3$.
These notations for the exponents $\nu_G$, $\nu_D$ at fixed values
of either $G$ or $D$ are akin to those traditionally used in thermodynamics.

For the poor solvent a similar procedure gives the 
following values for $N > 200$,
\begin{eqnarray}
\mbox{GSC:}\ & A_G = 0.560 \pm 0.007,\qquad  & \upsilon_G = 0.331\pm 0.002
   \qquad \mbox{for $U^{(0)} = 7$},\\
\mbox{GSC:}\ & A_G = 0.516 \pm 0.004,\qquad  & \upsilon_G = 0.333\pm 0.001
   \qquad \mbox{for $U^{(0)} = 10$},\\
\mbox{MC:}\ & A_G = 0.60 \pm 0.004,\qquad  & \upsilon_G = 0.31\pm 0.02
   \qquad \mbox{for $U^{(0)} = 6$},
\end{eqnarray}
or, roughly speaking, 
$\upsilon_G \approx \upsilon = \frac{1}{3}$,
with coefficient $A_G$ decreasing slowly with increasing $U^{(0)}$
and being almost independent of other parameters such as $G$.
Hence, we have for a poor solvent,
\begin{equation} \label{RgUps}
\sqrt{3 {\cal R}_g^2} \sim N^{\upsilon} \sim D^{\upsilon_G}\,(F-1)^{\upsilon_D\,(G+1)},
\ \ \upsilon_D=\upsilon_G \simeq \upsilon=\frac{1}{3}.
\end{equation}

It is interesting to note that for the ideal chain
(absence of the excluded volume interactions, $d^{(0)}=0$) we obtain
the following accurate asymptotic expression \cite{Comment}
for large $D$ and $G$,
\begin{equation}
\label{RgIdeal}
\sqrt{3{\cal R}_g^2} \simeq \ell\,\sqrt{3}\,D^{1/2}\,G^{1/2}.
\end{equation}
Therefore, generally, the dependence of the radius of gyration on the
spacer exponent is always the same as for an open/ring chain, but
the ideal system lacks the complicated dependence on the number of
generations $G$. It is worthwhile to comment also that such ideal
system should not be confused with the system in the
theta--solvent near the transition point $U^{(0)}_{{\cal E}'}$.
For the latter we shall have a scaling akin to
Eqs.~(\ref{RgNu},\ref{RgUps}),
\begin{equation} \label{RgTheta}
\sqrt{3 {\cal R}_g^2} \sim D^{{\sf v}_G}\,(F-1)^{{\sf v}_D\,(G+1)},
\ \ {\sf v}_G=\frac{1}{2}, \ \ \upsilon_D \le {\sf v}_D \le \nu_D.
\end{equation}
The values for ${\sf v}_G$, ${\sf v}_D$ obtained from the GSC theory
are reported in  Tabs.~\ref{tab:1},\ref{tab:2}.
We can conclude that indeed ${\sf v}_G=1/2$ with a high accuracy,
whereas ${\sf v}_D\simeq 0.35$. We expect the latter to be somewhat
lower in the MC simulation than in the GSC theory again.

\section{Discussion and Conclusion}\label{sec:concl}

In this paper we have studied a range of homo-- dendrimers
with $D=1-6$ spacers and $G=1-7$ generations in the good and poor
solvents, as well as across the coil--to--globule transition.
These studies have been performed by two different techniques
yielding a satisfactory agreement between their results. 
The first technique
is based on the Gaussian self--consistent (GSC) method, while the
second one is that of Monte Carlo (MC) simulation in continuous space.
Both rely on the same coarse--grained model involving harmonic
springs for connected monomers and the Lennard--Jones pair--wise
potential with a hard core part.
The overall accord between these two rather distinct approaches
is quite important due to the limitations of both techniques.
The MC technique, while in principle exactly modelling a given system,
in practice is always limited by the statistical accuracy and quality
of equilibration for very large dendrimers computationally affordable.
The GSC method, on the other hand, is about $10^3$ faster for the
studied range of dendrimer sizes and it directly produces the 
ensemble averaged equilibrium observables.
However, it is only an approximate technique, which has some
intrinsic inaccuracies, for instance somewhat overestimating the size of
polymers in the good solvent.

We have proposed a new scaling law for the radius of gyration of dendrimers
which involves the two exponents $\nu_G$ (at fixed $G$) and
$\nu_D$ (at fixed $D$),
\begin{equation}
\label{ScalingConcl}
{\cal R}_g \sim D^{\nu_G}\,(F-1)^{\nu_D(G+1)} \sim D^{\nu_G-\nu_D}\,N^{\nu_D},
\end{equation}
where $\nu_G = \nu$, with $\nu\approx 3/5$, $1/2$, and $1/3$
being the inverse fractal dimension of an open/ring chain in the
good, ideal and poor solvents respectively.
While this agrees with the conclusions of Refs.~\onlinecite{ChenCui,Biswas}
for $\nu_G$, we have obtained a rather different dependence
on the number of generations.
Namely, we have concluded that 
$\nu_D \approx \nu-1/4 \simeq 0.338$ for the good solvent, 
and $\nu_D \approx \nu=1/3$ for the globule. This is consistent
with Ref. \onlinecite{MuratGrest}, but our estimate of $\nu_D$
for the good solvent is more accurate enabling us to claim
that $\nu_D$ is certainly above, although quite close to
the poor solvent value $1/3$. 

It is important to emphasise also that the law in Eq. (\ref{ScalingConcl})
is valid for the model with the harmonic bonds employed here. Dendrimers with
bonds possessing a finite stretchability and angular potentials would
in all likelihood have more complicated and non--universal features
in their behaviour.
However, the harmonic model is an important and well tractable 
reference point for
further studies at the level of full atomistic detail for concrete
dendrimers, which could have bonds stretchable in different ranges.

We have concluded that the coil--to--globule 
transition for dendrimers is continuous
and that for the theta--solvent the scaling law
for ${\cal R}_g$ differs from that of the ideal dendrimer, in which
${\cal R}_g\sim (DG)^{1/2}$, having instead the $G$-dependence
of the form given by Eq.~(\ref{ScalingConcl}) 
with $1/3 \le \nu_D \le 3/5-1/4$.

Details of the conformational structure of dendrimers 
have been analysed for the good and poor solvents also. 
In some respects, our results support
certain observations of Ref.~\onlinecite{MansfieldKlushinMacromol}
made for the good solvent. Namely, we find a strong evidence of the
re--entry of the terminal generation into the dendrimer up to its very
centre, producing a rather delocalised partial density for those
monomers. This results in the highest total density near the 
core for relatively small dendrimers, 
which decreases in larger dendrimers due to the
`entropic pull' effect of extra branches.
However, as the number of generations exceeds $G=5$
there also develops a lower density domain inside the dendrimer.
Moreover, we have analysed the bond stretching along a continuous branch.
This is highest near the core, especially for dendrimers with a
large number of generations. The stretching decreases 
with increasing generation index and oscillates in the magnitude 
with the spacer index. This observation is interpreted consistently
with the connectivity contribution to the mean energy.

Furthermore, we have attempted to elucidate the degree of the steric congestion 
in a dendrimer by considering the shell index dependence of the 
monomer packing coefficient $\eta_{\rho}$. 
The latter is the volume ratio of the space occupied
by all other monomers around a given one.
In a relatively small dendrimer $\eta_{\rho}$ oscillates in the
spacer index with only a weak generation index dependence, except
for the terminal generation which is less congested.
A dendrimer with a large number of generations, however, has 
a less congested hollow domain inside it and the highest congestion in
the penultimate generations in accord with the density distributions.

The degree of asphericity of dendrimers appears to decrease with
the increasing number of generations, depending rather weakly on
the number of spacers. This is also confirmed by the behaviour of
the static structure factor, which develops an oscillating globule--like
character for larger $G$.

Interestingly, while incorrect in concrete detail, the onion model predictions 
of Ref.~\onlinecite{deGennesHervet} do have a rather limited meaning in a sense
that there exists a hollow domain inside a dendrimer of large enough
number of generations, but not at the core itself.
However, the partial densities of generations fall off only beyond
the appropriate cut--off radii. These densities are
totally delocalised until the cut--offs
(see Fig~\ref{fig:DensD2G}b),
being fairly large at the core \cite{MuratGrest},
and certainly are not concentrated within thin layers as surmised.

A similar analysis has been carried out for the poor solvent as
well. The globular conformation of a dendrimer is much more compact,
and it is rather uniform without any hollow domains. The packing coefficient
for the globule is much larger in value than for the good solvent and depends
on the spacer index only for the monomers in the inner generations.
There is also much less stretching of bonds involved near the core
than for the good solvent, even in a dendrimer with a large $G$.
Generally, the bond stretching in the globule is stronger for the
branch points than for the middle of spacers or terminal monomers.
Thus, the connectivity of a dendrimer in the globular state
manifests itself in a much more localised manner, and with a
significantly smaller effect than for the good solvent.

We believe that our results allow one to reconcile some of the 
disparate predictions from previous theoretical and simulation 
studies by finding some simple limiting regimes and 
by discovering non--monotonic dependencies of various observables
and the local structure of dendrimers on the number of generations.
These unusual features of dendrimers indeed make them extremely
appealing for numerous future applications ranging from catalysis,
rheology modifiers, to drug delivery and biotechnology more generally.

\acknowledgments

The authors acknowledge interesting discussions with Professor
G.~Allegra, Dr G.~Raos, and particularly with
Professor F.~Ganazzoli who attracted our attention to the field of dendrimers.
We are also grateful to Gillian Flanagan and Marese O'Brien
for their help and insight during their undergraduate research projects.
Support from the IRCSET basic research grant SC/02/226
is also acknowledged.

\appendix

\section{The topological tree}
\label{app:A}

The concept of the kinematic symmetries can be best explained by using
the topological tree presented in Fig. \ref{fig:eqtree},
which is equivalent to the dendrimer diagram of Fig. \ref{fig:dendri}.
Hence, the horizontal line would represent any selected consecutive branch
of the dendrimer, which we shall name the primary branch ${\cal B}$,
with the core monomer being on the left. The length of ${\cal B}$
is then $n_{{\cal B}}=D(G+1)$, excluding the core monomer itself.
Any two branches become topologically non--equivalent
to each other as long as they diverge from the common parent
sub-branch at some value of the generation index $g$.
We shall call any two such branches non--equivalent of order $g$.
The leftmost vertical line on the left of Fig. \ref{fig:eqtree}
represents any branch non--equivalent
to ${\cal B}$ of order $g=0$, which we shall denote as
${\cal B}_{0}$ with the length $n_{{\cal B}_0}\equiv n_0$.
Similarly then, we shall have branches ${\cal B}_g$ non--equivalent
to ${\cal B}$ of order $g$ with the length $n_g=D(G-g+1)$ for all
values of $g=0,1,\ldots G$ in our topological tree.
Any two pairs of monomers then would have ${\cal D}_{ij}={\cal D}_{i'j'}$
as long as the pair $(i',j')$ lies in the same places in the
topological tree as the pair $(i,j)$, thereby defining the equivalence
classes of monomer pairs.

Moreover, we can easily count the number of independent degrees
of freedom in the matrix ${\cal D}_{ij}$. For that we have to
count the number of possible pairs $(\rho,\rho')$ in Fig. \ref{fig:eqtree}
such that e.g. $\rho' \leq \rho$, where $i=(\rho,\varphi)$ and
$i'=(\rho',\varphi')$. Accidentally, we note that the 
topological tree 
disregards the angular coordinates $\varphi$ of any equivalent
monomers. 
There are two possibilities of pairing $\rho$ and $\rho'$:
a) $\rho,\rho' \in {\cal B}$ and b) $\rho \in {\cal B}$ and
$\rho' \in {\cal B}_g$ for some $g$. In the former case we get the
number of pairs $C_{{\cal B}}=\sum_{i=1}^{n_{\cal B}}i = n_{\cal B}
(n_{\cal B}+1)/2$, 
whereas in the latter case we obtain $C_g=\sum_{i=n_g}^{1}i=n_g (n_g+1)/2$.
Thus, we have derived the total number of independent degrees of freedom,
\begin{equation}
\label{degInd}
C_{ind}^{(2)}=C_{{\cal B}}+\sum_{g=G}^{0} C_g=
(G+1)\frac{D}{2}\left(D\left(\frac{G^2}{3}+\frac{13}{6}G+2
\right)+2+\frac{G}{2}\right),
\end{equation}
which is identical to Eq. (A-1) in Ref. \onlinecite{Ganazzoli-DendriTerra}.

Similarly, the total number of independent 1-point observables
(e.g. the packing coefficient $\eta_i$),
which depend on $\rho$ only, 
is simply the length of the primary branch ${\cal B}_0$,
\begin{equation}
C^{(1)}_{ind} = D(G+1)+1.
\end{equation}





\newpage
\section*{Figure Captions}

\begin{figure}
\caption{ \label{fig:Snap}
Typical snapshots of conformations of the dendrimers
D3G3 (Fig.~a) and D3G7 (Fig.~b) in the good solvent from MC simulation.
The core monomer is represented as a sphere of the monomer real
diameter $d^{(0)}=\ell$ in both figures.
}
\end{figure}

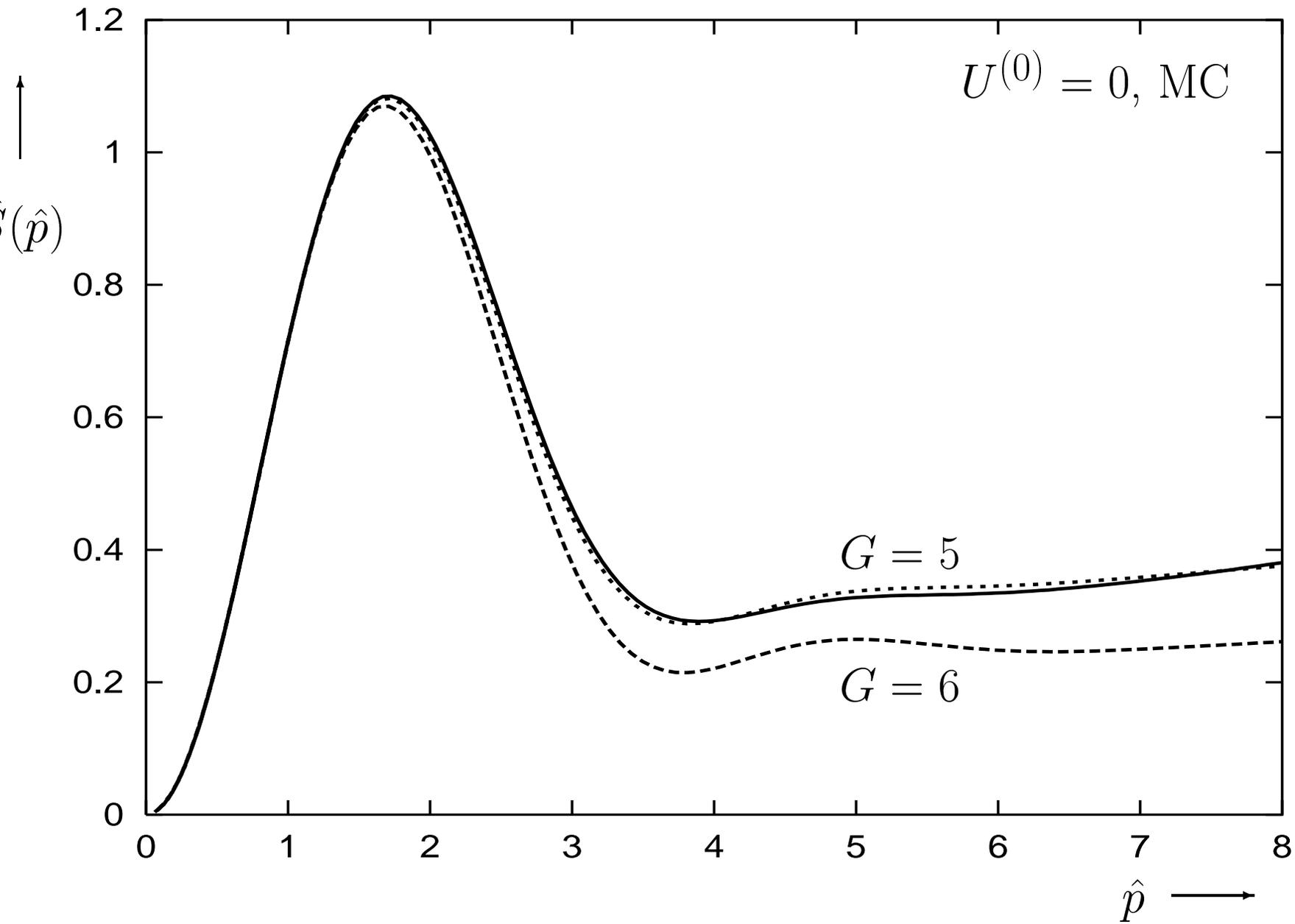
\begin{figure}
\caption{ \label{fig:Kratky}
Kratky plots of the rescaled static structure factor 
$\hat{S}=p^2\,3{\cal R}_g^2\,S(p)/N$ vs the rescaled
wave number $\hat{p}=p\,\sqrt{3{\cal R}_g^2}$ for
dendrimers in the good solvent ($U^{(0)}=0$) from MC simulation.
Here the solid line corresponds to D1G5, the
long--dashed curve --- to D1G6 and the short--dashed
curve --- to D2G5 dendrimer.
The largest relative statistical errors
in this plot occur for the D2G5 dendrimer and are equal to:
$\delta\hat{p} = 0.1\%$ and
$\delta\hat{S} \leq 0.3\%$.
}
\end{figure}

\begin{figure}
\caption{ \label{fig:DensD2G}
Plots of the monomer densities $g^{(1)}_g$ vs the separation
from the centre--of--mass ${\bf r}_{cm}$ for the dendrimers D2G4 
(Fig.~a) and D2G6 (Fig.~b) in the good solvent 
($U^{(0)}=0$) from MC simulation.
Insets of figures show the monomer densities
of the same dendrimers in the poor solvent ($U^{(0)}=6$).
Curves from top to bottom as on the left--hand side correspond to
the following densities.
In main bodies of figures:
the total density $\sum_g g^{(1)}_g$ (solid line), density of
non--terminal generations $\sum_{g<G} g^{(1)}_g$ (dashed line), 
density of the innermost generations $g^{(1)}_{-1}+g^{(1)}_{0}$
(thin solid line with quadrangles), 
density of the terminal generation $g^{(1)}_G$
(dotted line), and density of the penultimate
generation $g^{(1)}_{G-1}$ (thin line with triangles);
In insets of figures:
the total density $\sum_g g^{(1)}_g$ (solid line), density of
non--terminal generations $\sum_{g<G} g^{(1)}_g$ (dashed line), 
density of the terminal generation $g^{(1)}_G$
(dotted line).
Errorbars represent statistical errors of the data from MC simulation.
}
\end{figure}

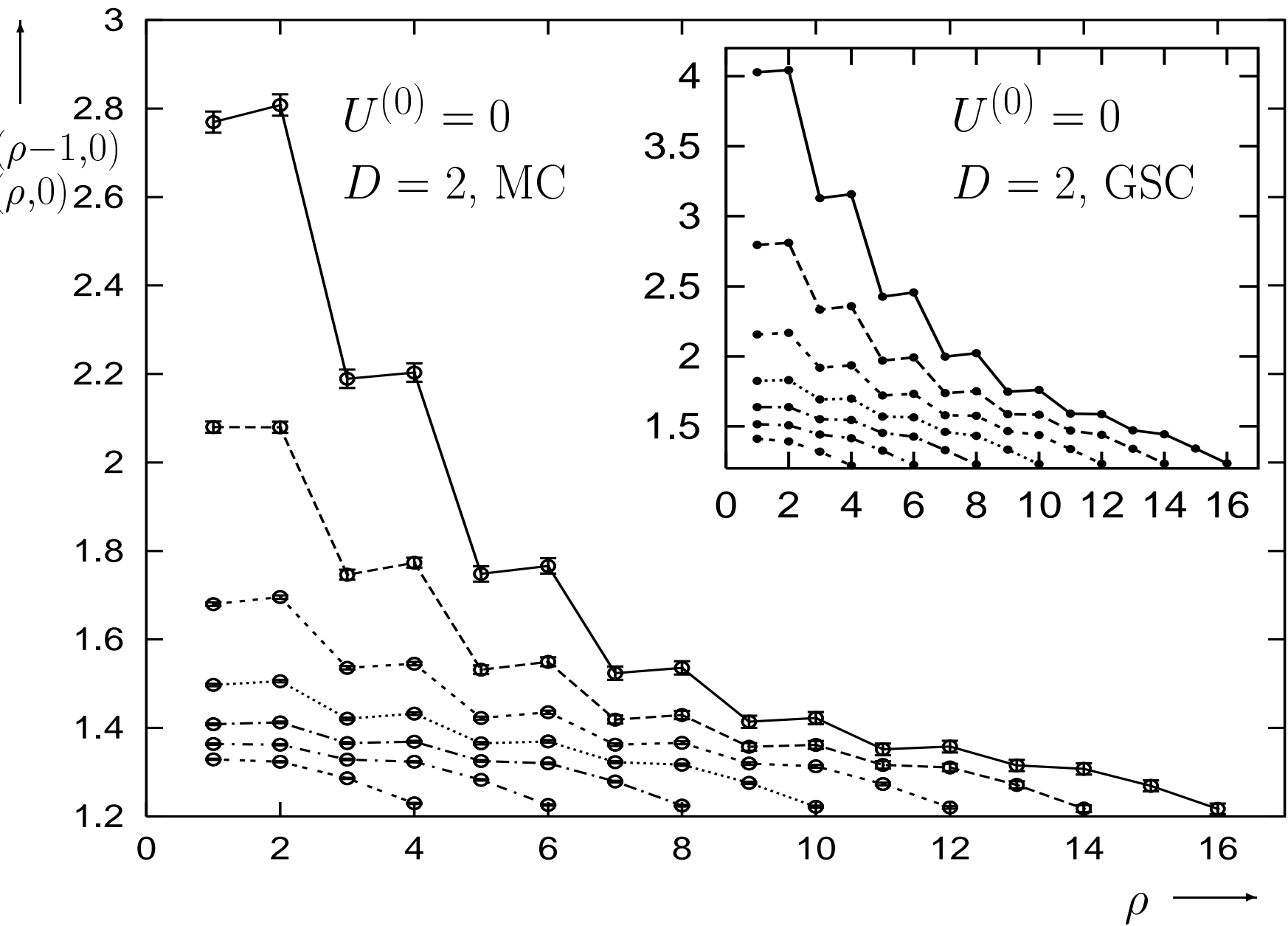
\begin{figure}
\caption{ \label{fig:DrrCoilD2}
Plots of the mean--squared distances $D_{(\rho-1,0)\,(\rho,0)}$
between radial nearest neighbours vs the shell index $\rho$
from MC simulation (main body of the figure)
and from the GSC theory (inset of the figure)
for dendrimers with $D=2$ spacers in the good solvent ($U^{(0)}=0$).
The curves correspond to $G=1,2,3,4,5,6,7$ (from bottom to top).
Errorbars represent statistical errors of the data from MC simulation.
}
\end{figure}

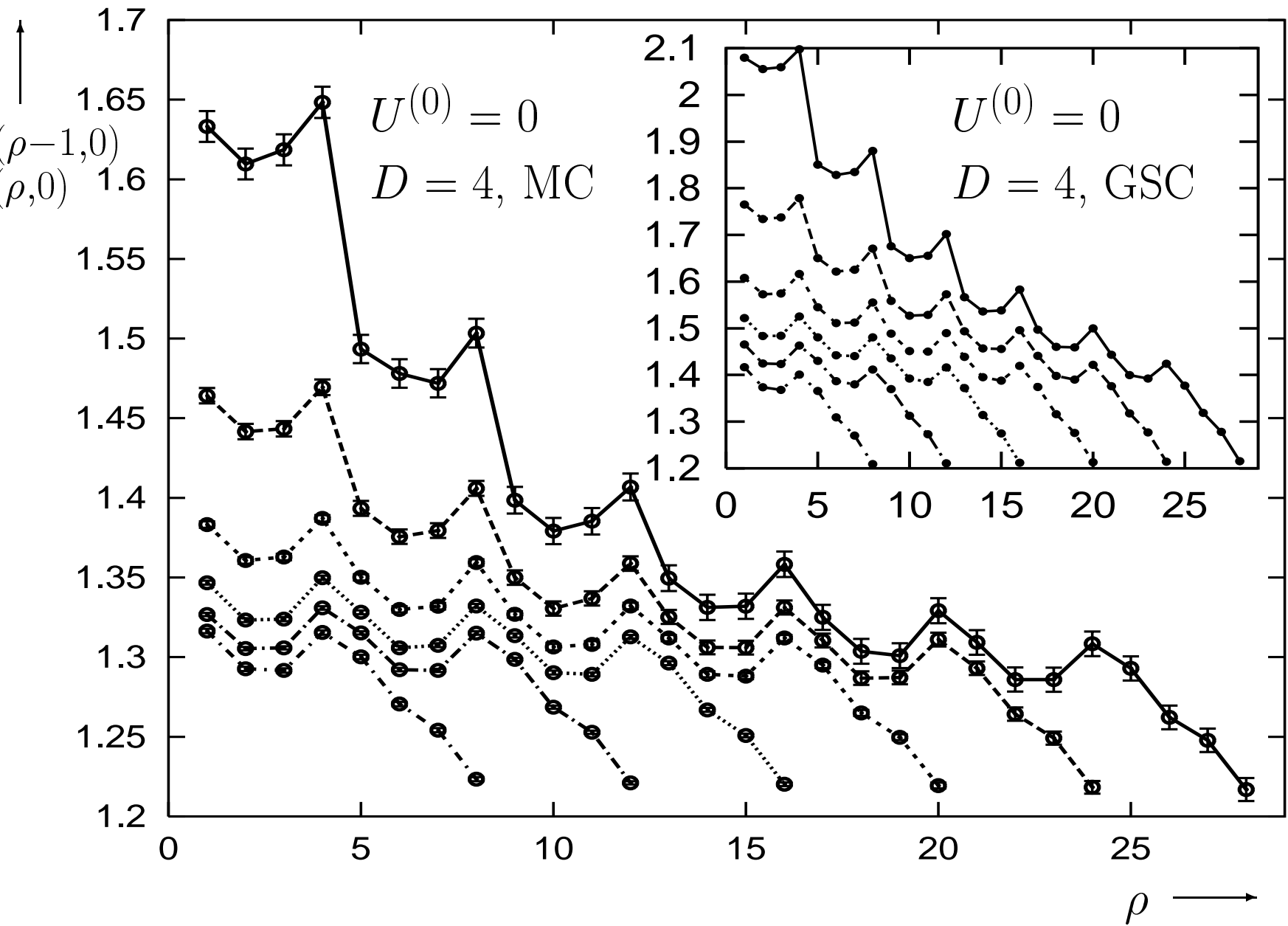
\begin{figure}
\caption{ \label{fig:DrrCoilD4}
Plots of the mean--squared distances $D_{(\rho-1,0)\,(\rho,0)}$
vs the shell index $\rho$ from
MC simulation (main body of the figure)
and from the GSC theory (inset of the figure)
for dendrimers with $D=4$ spacers in the good solvent ($U^{(0)}=0$).
The curves correspond to $G=1,2,3,4,5,6$ (from bottom to top).
Errorbars represent statistical errors of the data from MC simulation.
}
\end{figure}

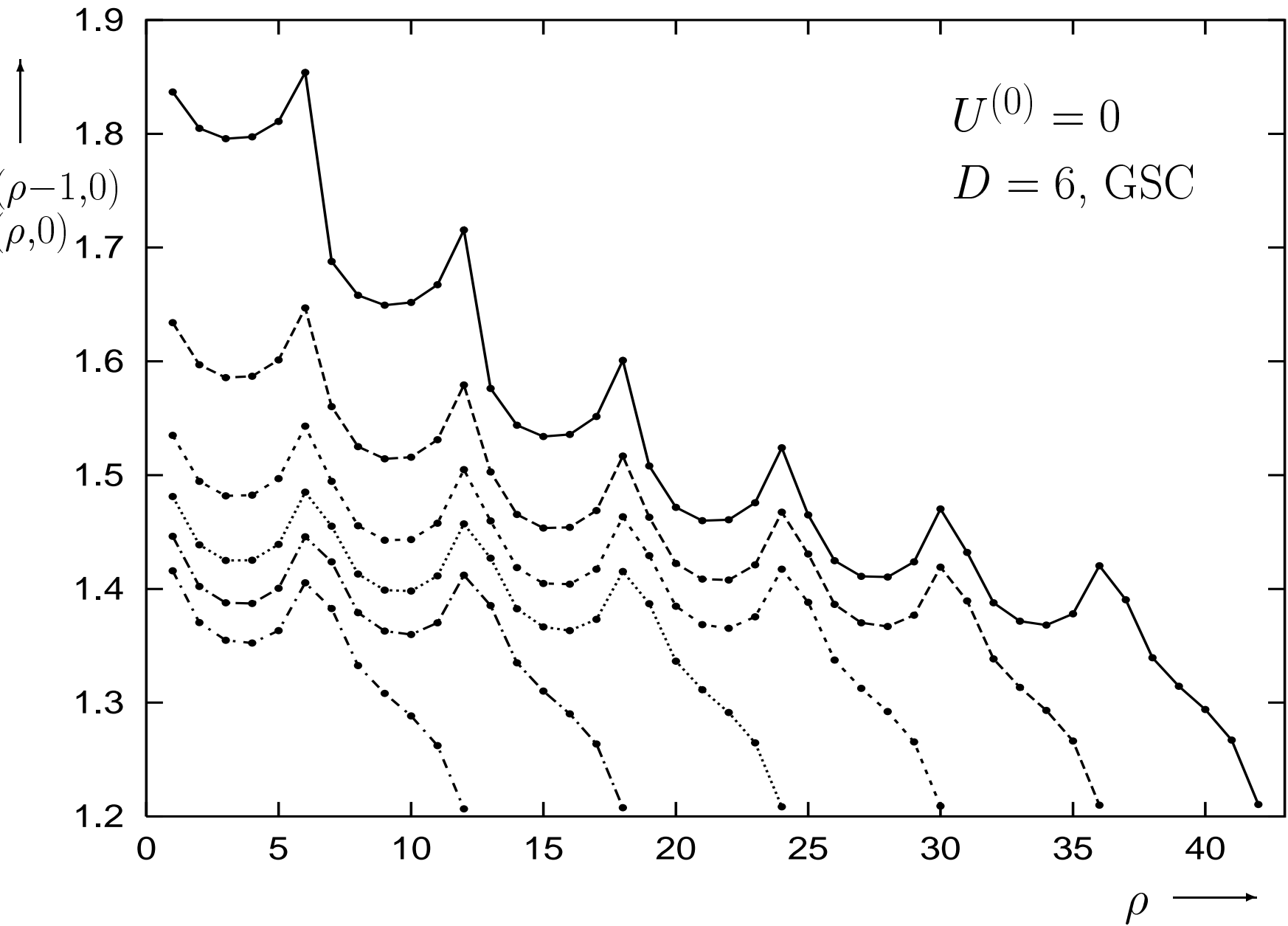
\begin{figure}
\caption{ \label{fig:DrrCoilD6}
Plots of the mean--squared distances $D_{(\rho-1,0)\,(\rho,0)}$
vs the shell index $\rho$ from the GSC theory
for dendrimers with $D=6$ spacers in the good solvent ($U^{(0)}=0$).
The curves correspond to $G=1,2,3,4,5,6$ (from bottom to top).
}
\end{figure}

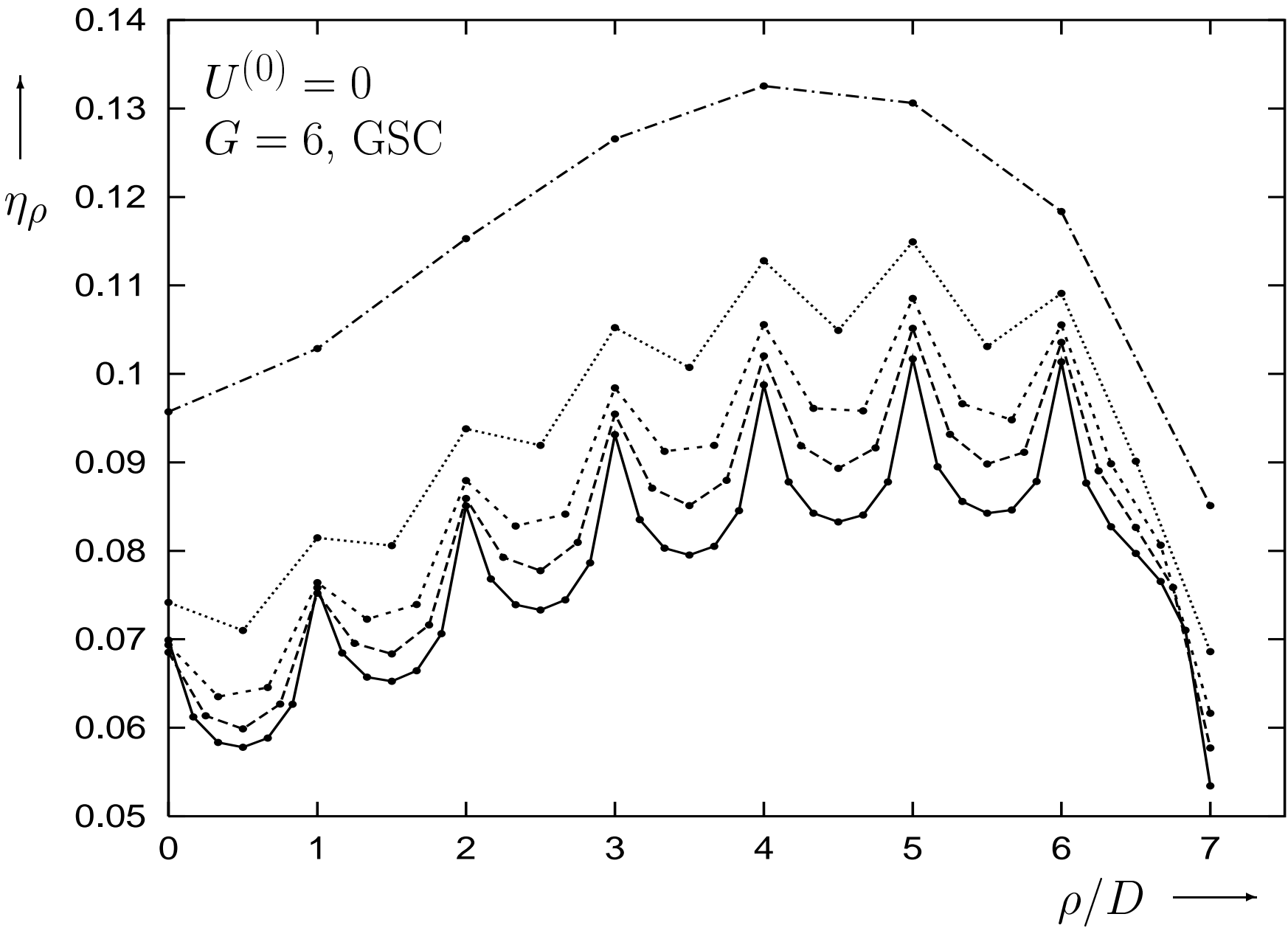
\begin{figure}
\caption{ \label{fig:EtaCoilG6}
Plots of the packing coefficient $\eta_{\rho}$ defined by
Eqs.~(\ref{eq:etai},\ref{eq:Fy}) vs the rescaled shell index $\rho/D$
for dendrimers consisting of $G=6$ generations in the good solvent
($U^{(0)}=0$) from the GSC theory. 
The curves correspond to the following values of
$D=1,2,3,4,6$ (from top to bottom).
}
\end{figure}

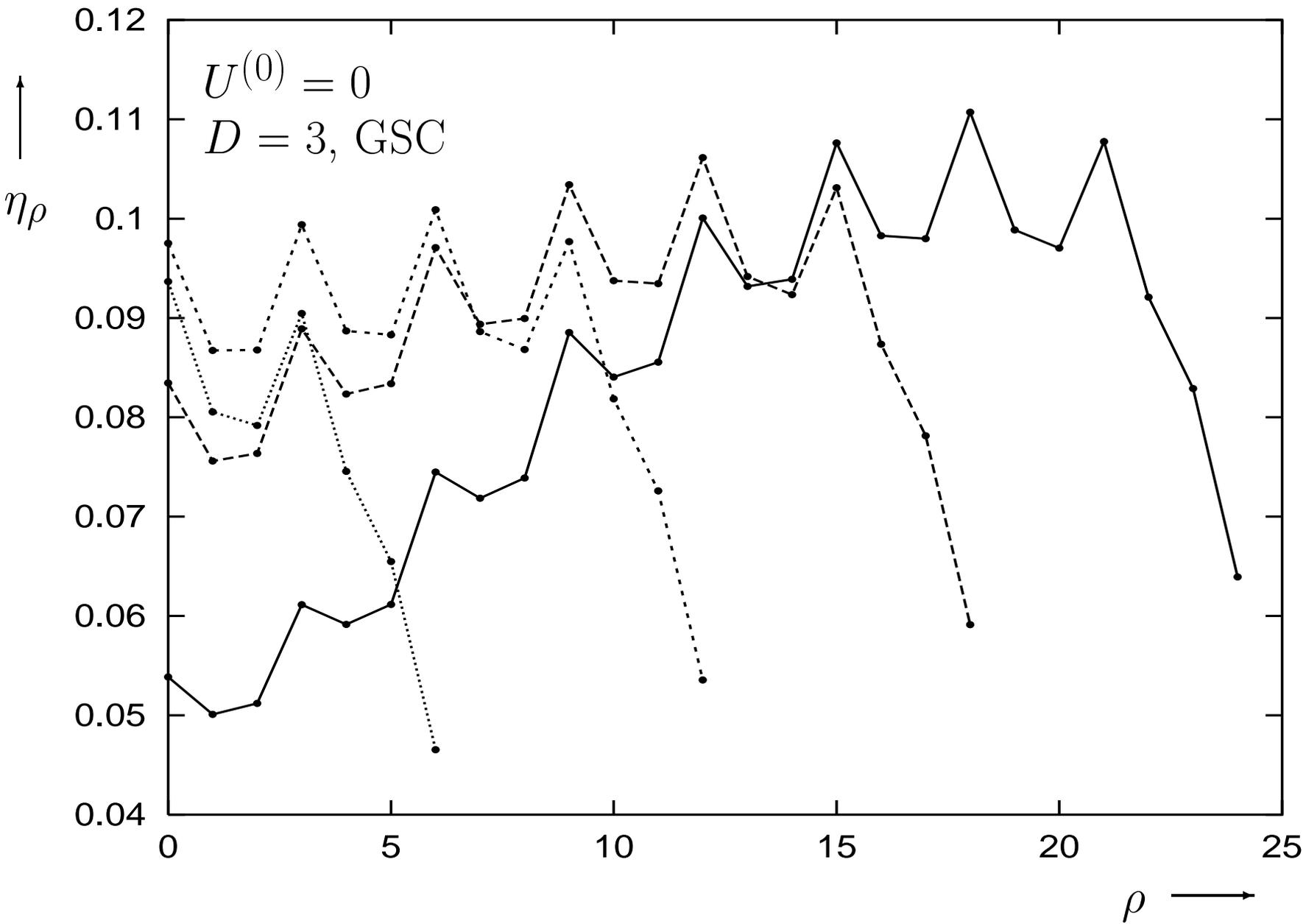
\begin{figure}
\caption{ \label{fig:EtaCoilD3}
Plots of the packing coefficient $\eta_{\rho}$ 
vs the shell index $\rho$
for dendrimers with $D=3$ spacers in the good solvent
($U^{(0)}=0$) from the GSC theory. 
The curves correspond to the following values of $G$:
$G=1$ (dotted line), $G=3$ (short--dashed line), $G=5$ (long--dashed line),
and $G=7$ (solid line).
}
\end{figure}

\begin{figure}
\caption{ \label{fig:RgColTrans}
Plots of the mean--squared radius of gyration
$3{\cal R}_g^2$ (Fig.~a) and the specific energy slope
$N^{-1}d{\cal E}/dU^{(0)}$
(Fig.~b) vs the degree of Lennard--Jones attraction $U^{(0)}$
across the coil--to--globule transition from the GSC theory 
for dendrimers with $G=5$ generations
and $D=1,2,3,4,6$ (from bottom to top as on the left--hand side).
}
\end{figure}

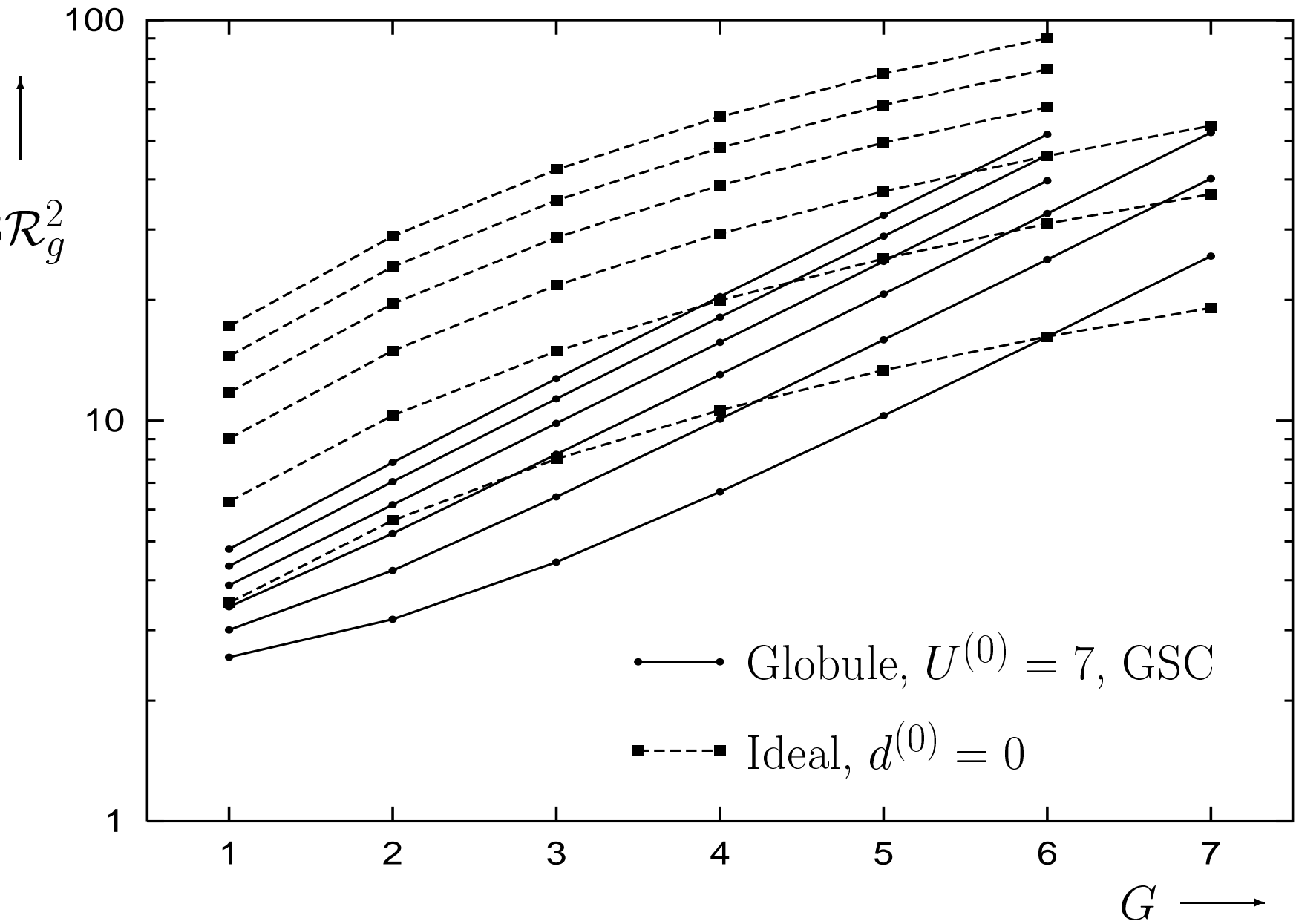
\begin{figure}
\caption{ \label{fig:RgGloblG}
Plots of the mean--squared radius of gyration
$3{\cal R}_g^2$ vs the number of generations $G$
for the dendrimers in the poor solvent, $U^{(0)}=7$, from the GSC theory
(filled circles and solid lines)
and for the ideal dendrimers, $d^{(0)}=0$
(filled quadrangles and dashed lines).
The curves correspond to the following values of $D$
for each of the two families: $D=1,2,3,4,5,6$ (from bottom to top).
}
\end{figure}

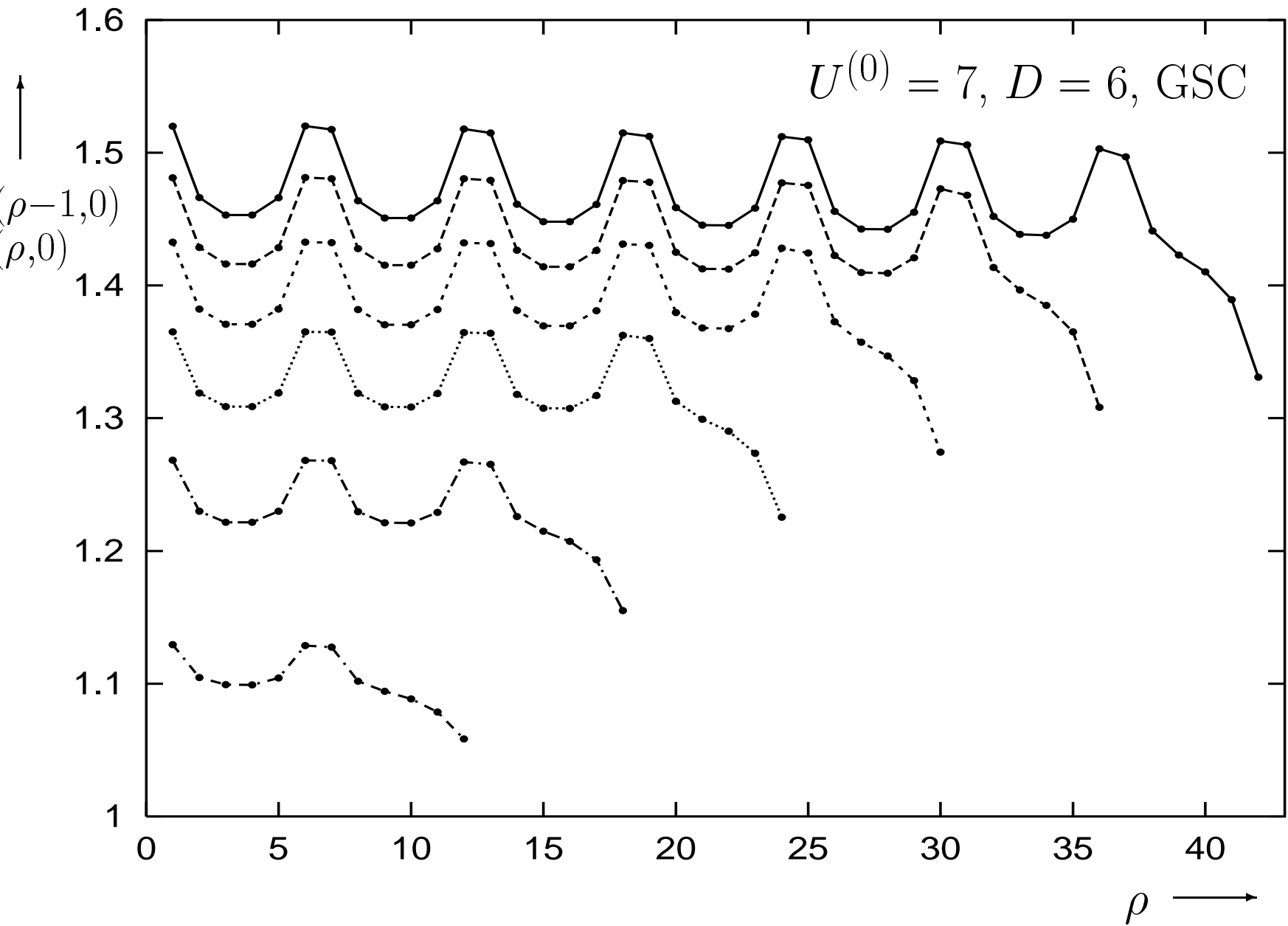
\begin{figure}
\caption{ \label{fig:DrrGlobD6}
Plots of the mean--squared distances $D_{(\rho-1,0)\,(\rho,0)}$
vs the shell index $\rho$ from the GSC theory
for dendrimers with $D=6$ spacers in the poor solvent ($U^{(0)}=7$).
The curves correspond to $G=1,2,3,4,5,6$ (from bottom to top).
}
\end{figure}

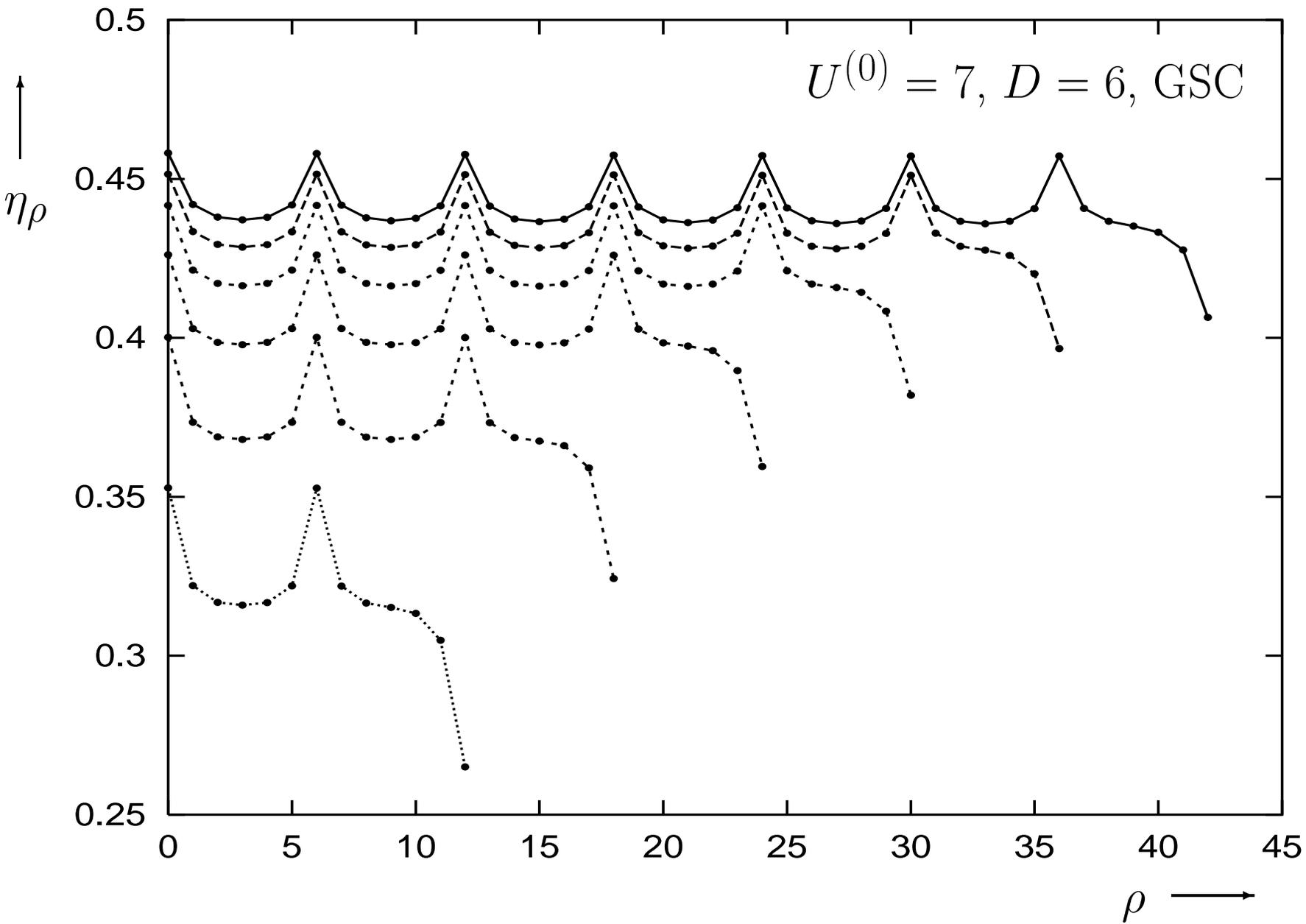
\begin{figure}
\caption{ \label{fig:EtaGlobD6}
Plots of the packing coefficient $\eta_{\rho}$ 
vs the shell index $\rho$
for dendrimers with $D=6$ spacers in the poor solvent
($U^{(0)}=7$) from the GSC theory. 
The curves correspond to the following values of
$G=1,2,3,4,5,6$ (from bottom to top).
}
\end{figure}


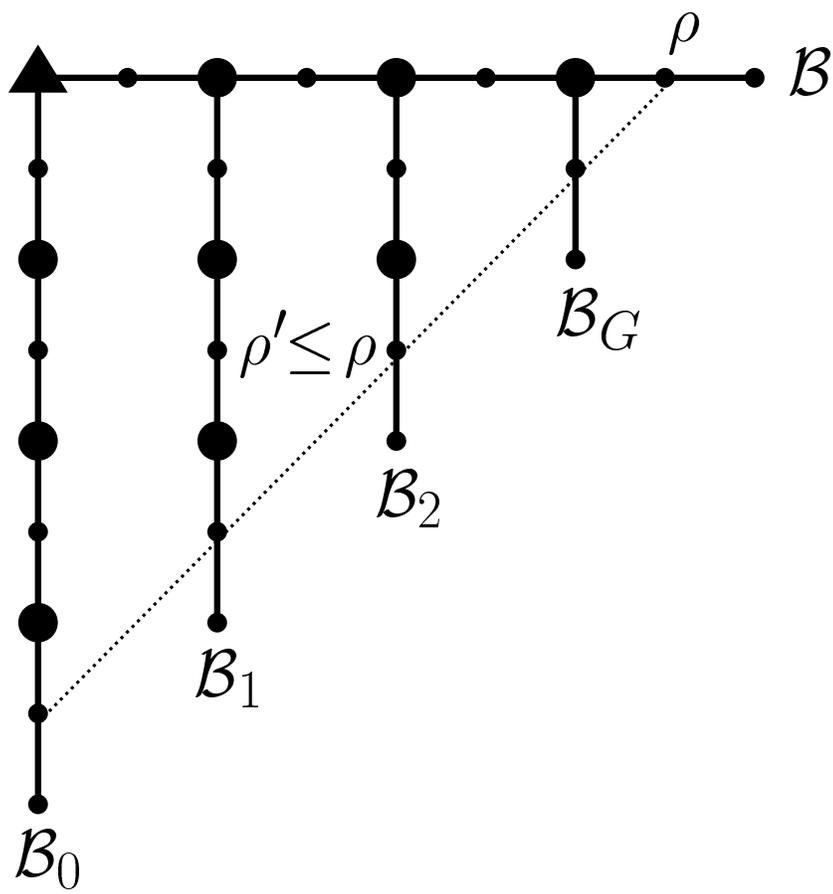
\begin{figure}
\caption{ \label{fig:eqtree}
Tree diagram topologically equivalent to that of
Fig.~\ref{fig:dendri} for D2G3 dendrimer.
Dotted line illustrates the area satisfying
condition $\rho' \leq \rho$. 
}
\end{figure}

\newpage

\section*{Tables}

\begin{table}
\caption{\label{tab:3}
Values of the pair--wise interaction parameter at the
coil--to--globule transition point
$U^{(0)}_{{\cal E}'}$ and at the special point $U^{(0)}_{{\cal R}}$
in parentheses vs $D$ and $G$.
The former has been determined by a quadratic interpolation of
the maxima positions of $-d^2{\cal E}/dU^{(0)\,2}$. The latter
has been determined by the condition,
${\cal R}_g^2 (U^{(0)}_{{\cal R}}) = {\cal R}_g^2 ({\rm ideal})$.
}
\vskip 5mm

\begin{tabular}{|c|cccccc|}
      & $G=1$ & $G=2$ & $G=3$ & $G=4$ & $G=5$ & $G=6$ \\
\hline
$D=1$ & 4.86 (4.28) & 3.87 (3.81) & 3.27 (3.75) & 2.97 (4.02)
      & 2.75 (4.78) & 2.54 (6.97) \\
$D=2$ & 4.05 (3.50) & 3.24 (3.20) & 2.87 (3.17) & 2.70 (3.33)
      & 2.61 (3.77) & 2.50 (4.90) \\
$D=3$ & 3.56 (3.17) & 2.96 (2.95) & 2.68 (2.92) & 2.55 (3.03)
      & 2.48 (3.36) & 2.43 (4.14) \\
$D=4$ & 3.29 (2.99) & 2.80 (2.79) & 2.57 (2.77) & 2.45 (2.87)
      & 2.39 (3.13) & 2.36 (3.74) \\
$D=5$ & 3.11 (2.86) & 2.69 (2.70) & 2.49 (2.67) & 2.39 (2.75)
      & 2.35 (2.98) &  --- \\
$D=6$ & 2.98 (2.77) & 2.61 (2.62) & 2.43 (2.60) & 2.35 (2.68)
      & 2.31 (2.87) &  ---  \\
\end{tabular}
\end{table}

\begin{table}
\caption{\label{tab:1}
Values of parameters $a_G$, $\nu_G$ and
${\sf a}_G$, ${\sf v}_G$
obtained from fitting procedure via Eqs.~(\ref{eq:fitFlory},\ref{eq:fitTheta})
for the good athermal, $U^{(0)}=0$, and 
the theta, $U^{(0)}_{{\cal E}'}$, solvents
over various values of $D$ at a given value of $G$ from 
the GSC and MC methods.
}
\vskip 5mm

\begin{tabular}{|c|cc||cc||cc|}
$G$ & GSC:   $a_G$  &         $\nu_G$  &  MC: $a_G$      &   $\nu_G$        &
      GSC:   ${\sf a}_G$  &   ${\sf v}_G$   \\
\hline
1 & $0.532\pm 0.009$ & $0.615\pm 0.005$ & $0.63\pm 0.01$  & $0.55\pm 0.01$  &
    $0.60\pm 0.02$   & $0.477\pm 0.009$ \\
2 & $0.445\pm 0.007$ & $0.626\pm 0.004$ & $0.50\pm 0.03$  & $0.58\pm 0.02$ &
    $0.51\pm 0.02$   & $0.489\pm 0.006$ \\
3 & $0.371\pm 0.005$ & $0.629\pm 0.003$ & $0.44\pm 0.03$  & $0.57\pm 0.01$ &
    $0.44\pm 0.02$   & $0.500\pm 0.005$ \\
4 & $0.312\pm 0.004$ & $0.628\pm 0.002$ & $0.35\pm 0.01$  & $0.58\pm 0.01$ &
    $0.39\pm 0.01$   & $0.498\pm 0.004$ \\
5 & $0.266\pm 0.004$ & $0.624\pm 0.002$ & $0.29\pm 0.02$  & $0.59\pm 0.01$ &
    $0.34\pm 0.01$   & $0.501\pm 0.006$ \\
6 & $0.232\pm 0.003$ & $0.621\pm 0.002$ & $0.26\pm 0.01$  & $0.57\pm 0.01$ &
    $0.30\pm 0.01$   & $0.497\pm 0.005$ \\
\end{tabular}
\end{table}

\begin{table}
\caption{\label{tab:2}
Values of parameters $a_D$, $\nu_D$ and
${\sf a}_D$, ${\sf v}_D$
obtained from fitting procedure via Eqs.~(\ref{eq:fitFlory},\ref{eq:fitTheta})
for the good athermal, $U^{(0)}=0$, and 
the theta, $U^{(0)}_{{\cal E}'}$, solvents
over various values of $G$
at a given value of $D$ from the GSC and MC methods.
}
\vskip 5mm

\begin{tabular}{|c|cc||cc||cc|}
$D$ & GSC:   $a_D$ &         $\nu_D$  &  MC: $a_D$      &   $\nu_D$        &
      GSC:   ${\sf a}_D$  &   ${\sf v}_D$   \\
\hline
1 & $0.98\pm 0.01$ & $0.377\pm 0.002$ & $1.08\pm 0.01$  & $0.333\pm 0.002$ &
    $0.69\pm 0.03$   & $0.380\pm 0.006$ \\
2 & $1.18\pm 0.02$ & $0.373\pm 0.002$ & $1.29\pm 0.02$  & $0.329\pm 0.002$ &
    $0.85\pm 0.03$   & $0.355\pm 0.005$ \\
3 & $1.33\pm 0.03$ & $0.370\pm 0.003$ & $1.43\pm 0.03$  & $0.330\pm 0.003$ &
    $0.92\pm 0.01$   & $0.346\pm 0.003$ \\
4 & $1.42\pm 0.05$ & $0.370\pm 0.004$ & $1.50\pm 0.05$  & $0.332\pm 0.004$ &
    $0.96\pm 0.02$   & $0.346\pm 0.003$ \\
5 & $1.52\pm 0.06$ & $0.369\pm 0.005$ & $1.48\pm 0.09$  & $0.340\pm 0.007$ &
    $0.97\pm 0.04$   & $0.349\pm 0.006$ \\
6 & $1.60\pm 0.07$ & $0.368\pm 0.005$ & $1.62\pm 0.09$  & $0.335\pm 0.008$ &
    $1.00\pm 0.04$   & $0.351\pm 0.005$ \\
\end{tabular}
\end{table}


\begin{references}

\bibitem{AdvPolHult}
A.~Hult, M.~Johansson, E.~Malmstr\"{o}m.
{\it Adv. Polym. Sci.} {\bf 143} 1 (1999).

\bibitem{DendrimerReview1}
G.R.~Newkome, C.N.~Moorefield, F.~V\"{o}gtle.
{\it Dendritic Molecules}. Verlag-Chemie, Weinheim, Germany (1996).

\bibitem{Tomalia}
D.A.~Tomalia, A.N.~Naylor, W.A.~Goddard III.
{\it Angew. Chem. Int. Ed. Engl.} {\bf 29} 138 (1990).

\bibitem{Frechet}
J.M.J.~Fr\'{e}chet.
{\it Science} {\bf 263} 1710 (1994).

\bibitem{Bosman}
A.W.~Bosman, H.M.~Janssen, E.W.~Meijer. {\it Chem. Rev.}
{\bf 99} 1665 (1999).

\bibitem{DendrimerReviewCatalysis}
G.E.~Oosterom, J.N.H.~Reek, P.C.J.~Kamer, P.W.N.M.~van~Leuwen.
{\it Angew. Chem. Int. Ed.} {\bf 40} 1829 (2001).

\bibitem{Groot}
D.~de~Groot, B.F.M.~de~Waals, J.N.H.~Reek, A.P.H.J.~Schenning,
P.C.J.~Kamer, E.W.~Meijer, P.W.N.M.~van~Leuwen.
{\it J. Am. Chem. Soc.} {\bf 123} 8453 (2001).

\bibitem{Liu}
M.~Liu, J.M.J.~~Fr\'{e}chet. {\it Pharm. Sci. Technol. Today}
{\bf 2 (10)} 393 (1999).

\bibitem{GuestMol}
J.F.G.A.~Jansen, E.M.M.~de~Brabander-van~den~Berg,
E.W.~Meijer. {\it Science} {\bf 266} 1226 (1994).

\bibitem{LaFerla}
R.~La~Ferla. 
{\it J. Chem. Phys.} {\bf 106} 688 (1997).

\bibitem{AdvPolFreire}
J.J.~Freire.
{\it Adv. Polym. Sci.} {\bf 143} 34 (1999).

\bibitem{ExperDendri}
R.~Scherrenberg, B.~Coussens, P.~van~Vliet, G.~Edouard, {\it et al}.
{\it Macromol.} {\bf 31} 456 (1998);
F.~Mallamace, E.~Canetta, D.~Lombardo, A.~Mazzaglia, {\it et al}.
{\it Physica} {\bf A 304} 235 (2002).

\bibitem{deGennesHervet}
P.G.~de~Gennes, H.~Hervet.
{\it J. Phys. (Paris)} {\bf 44} L351 (1983).

\bibitem{BorisRubinstein}
D.~Boris, M.~Rubinstein. {\it Macromol.} {\bf 29} 7251 (1996).

\bibitem{Ganazzoli-CondMat}
F.~Ganazzoli.
{\it Cond. Matter Phys.} {\bf 5} 1 (2002).

\bibitem{ConfTra}
E.G. Timoshenko, Yu.A. Kuznetsov, K.A.~Dawson. 
{\it Phys. Rev.} {\bf E 57} 6801 (1998).

\bibitem{Edwards}
M.~Doi, S.F.~Edwards. {\it The Theory of Polymer Dynamics}. Oxford Science,
New York (1989).

\bibitem{CloizeauxBook} J. des Cloizeaux, G.~Jannink.
{\it Polymers in Solution.} 
Oxford Science Publ. (1990).

\bibitem{Ganazzoli-DendriTerra}
F.~Ganazzoli, R.~La Ferla, G.~Terragni.
{\it Macromol.} {\bf 33} 6611 (2000).

\bibitem{Ganazzoli-DendriDynamics}
F.~Ganazzoli, R.~La Ferla, G.~Raffaini.
{\it Macromol.} {\bf 34} 4222 (2001).   

\bibitem{Ganazzoli-DendriUnper}
F.~Ganazzoli, R.~La Ferla.
{\it J. Chem. Phys.} {\bf 113} 9288 (2000).

\bibitem{TomaliaMD}
A.N.~Naylor, W.A.~Goddard III, G.E.~Kiefer,
D.A.~Tomalia.
{\it J. Am. Chem. Soc.} {\bf 111} 2339 (1989).

\bibitem{Lescanec}
R.L.~Lescanec, M.~Muthukumar.
{\it Macromol.} {\bf 23} 2280 (1990).

\bibitem{MansfieldKlushinMacromol}
M.L.~Mansfield, L.I.~Klushin.
{\it Macromol.} {\bf 26} 4262 (1993).

\bibitem{ChenCui}
Zh.Yu.~Chen, Sh.-M.~Cui.
{\it Macromol.} {\bf 29} 7943 (1996).


\bibitem{Biswas}
P.~Biswas, B.J.~Cherayil. {\it J. Chem. Phys.} {\bf 100} 3201 (1994).

\bibitem{MuratGrest}
M.~Murat, G.S.~Grest. {\it Macromol.} {\bf 29} 1278 (1996).

\bibitem{Torus} 
Yu.A.~Kuznetsov, E.G.~Timoshenko.
{\it J. Chem. Phys.} {\bf 111}, 3744 (1999).

\bibitem{CopStar}
F.~Ganazzoli, Yu.A.~Kuznetsov, E.G.~Timoshenko.
{\it Macromol. Theory Simul.} {\bf 10} 325 (2001).

\bibitem{CorFunc}
E.G. Timoshenko, Yu.A. Kuznetsov, R. Connolly. 
{\it J. Chem. Phys.} {\bf 116} 3905 (2002).

\bibitem{Computers}
We have used the 28-CPU Beowulf cluster (SCOP) described
in detail at http://darkstar.ucd.ie/cluster.

\bibitem{GSC0}
E.G. Timoshenko, Yu.A. Kuznetsov.
Accepted for publication in {\it J. Chem. Phys.}
{\bf 117: 11} issue 15th of Sep. 
AIP ID 503235JCP (2002); E-print: cond-mat/0207204.

\bibitem{CombStar}
E.G. Timoshenko, Yu.A. Kuznetsov. 
{\it Colloids and Surfaces} {\bf A 190} 135 (2001).

\bibitem{AllenTild}
 M.P.~Allen and D.J.~Tildesley (Ed.),
 {\it Computer Simulation of Liquids}. Clarendon Press, Oxford (1987).

\bibitem{CarnahanStarling}
N.F.~Carnahan, K.E.~Starling.
{\it J. Chem. Phys.} {\bf 51} 635 (1969).
          

\bibitem{Comment}
For $d^{(0)}=0$ the GSC theory and MC simulations 
exactly agree with the analytic
formula for ${\cal R}^{2}_g$ by La Ferla \cite{LaFerla}.
Note that we do not have a factor of $3$ in the first
term of Eq.~(\ref{cmc:hamil}) and that
the mean--squared radius of gyration in our notation is 
the quantity $3{\cal R}_g^{2}$.
Both of the factors of 3 cancel out therefore, so that formally
${\cal R}_g^{2}({\rm our})={\cal R}_g^{2}({\rm La\ Ferla})$.


\end{references}
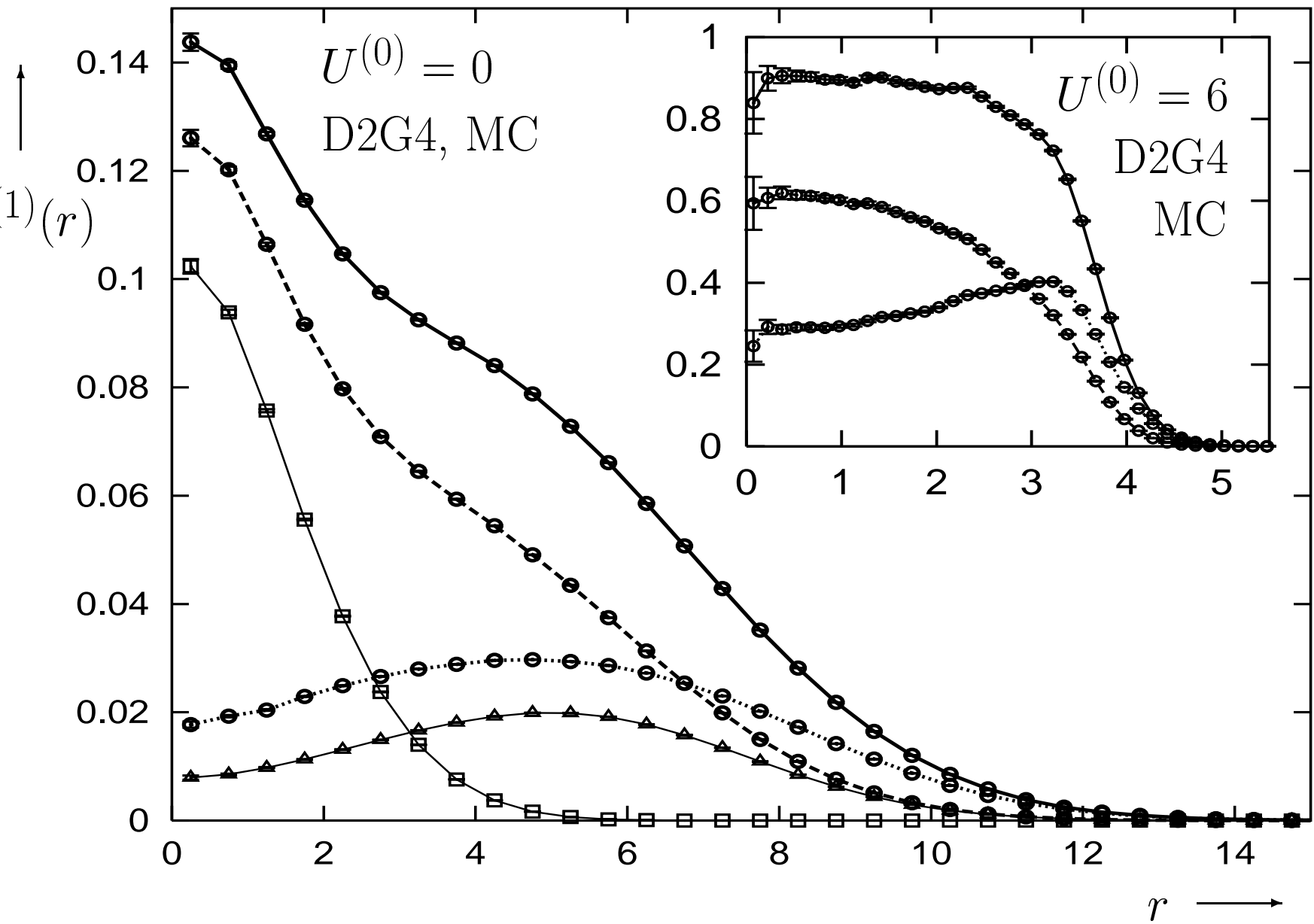
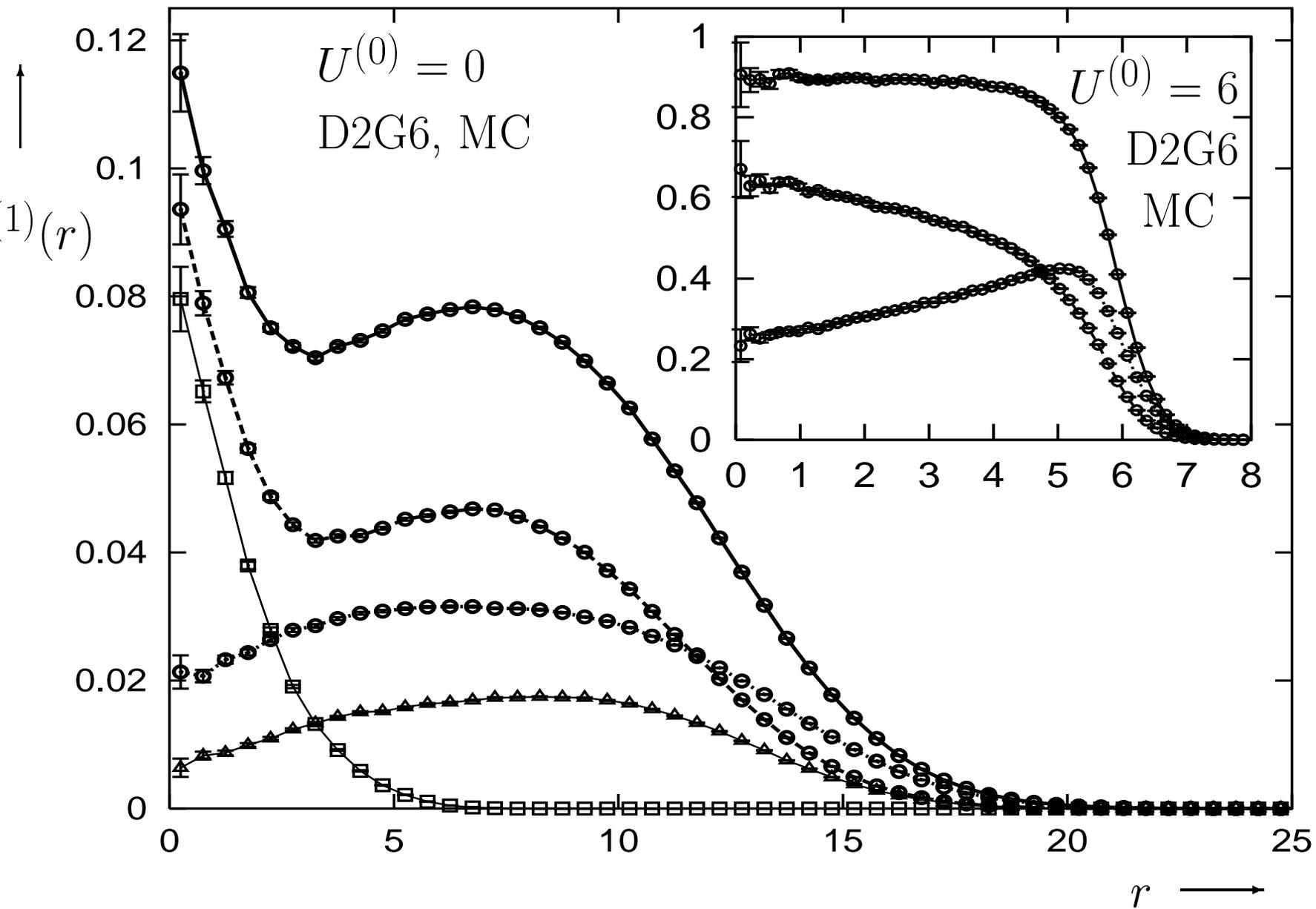
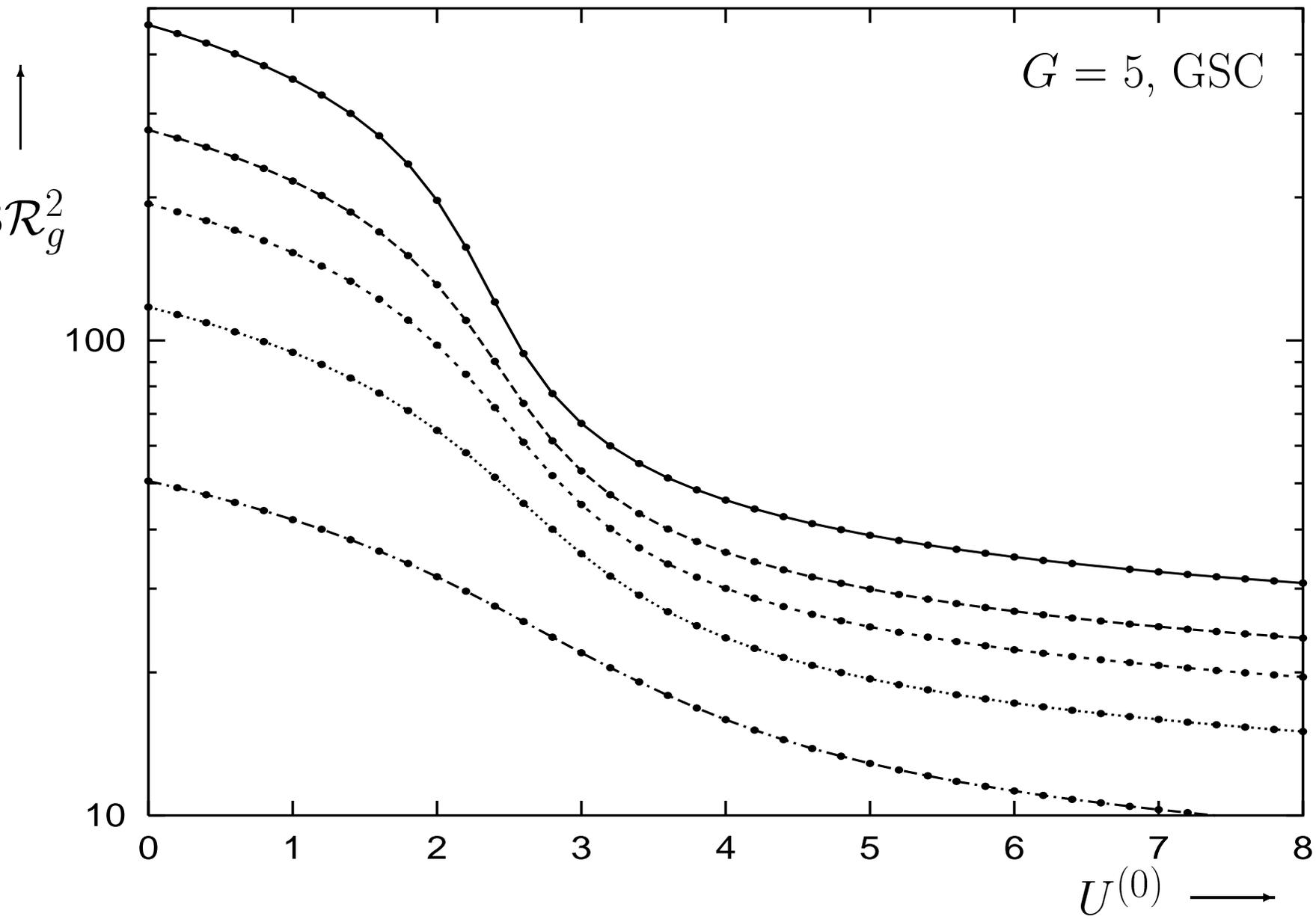
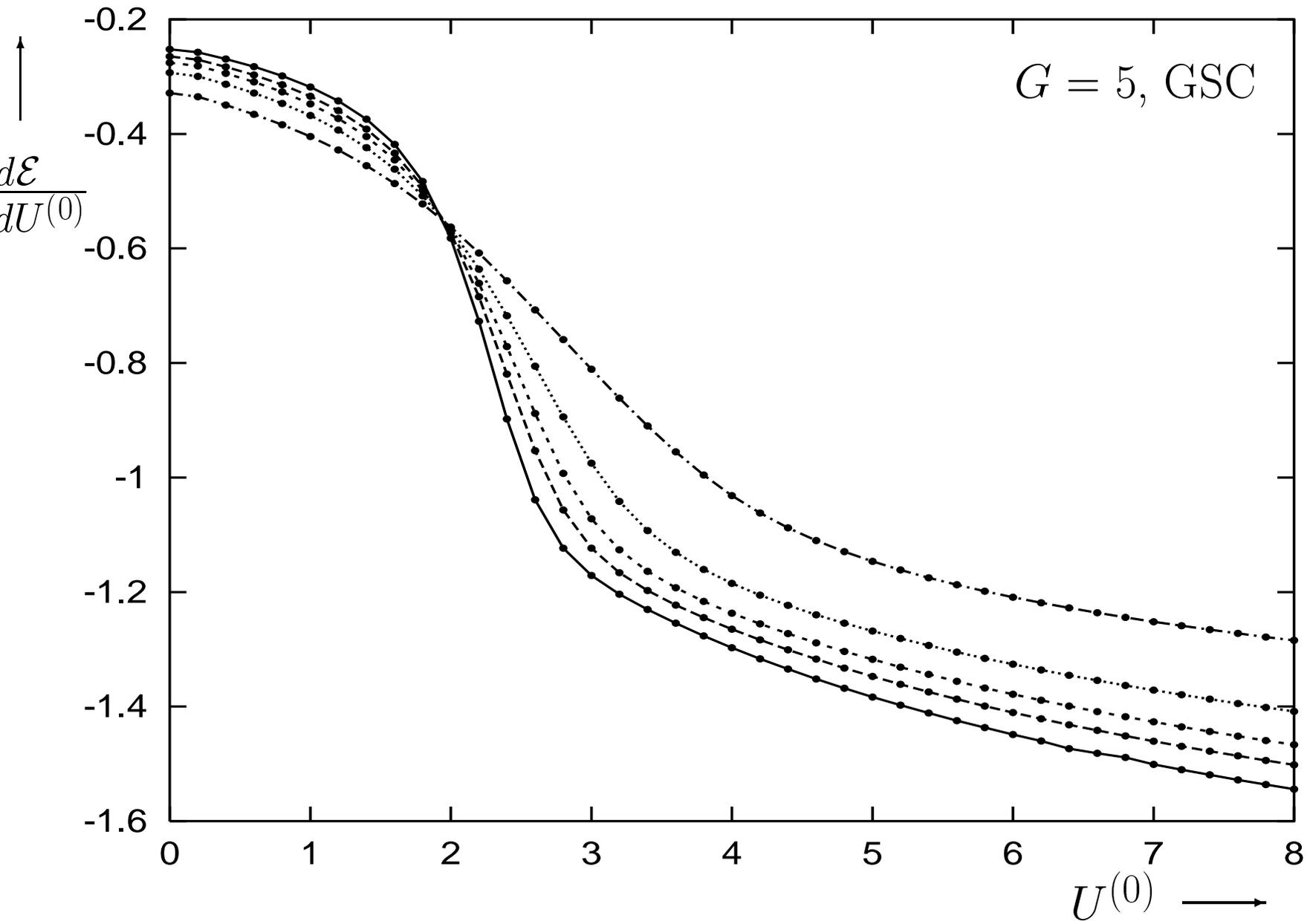
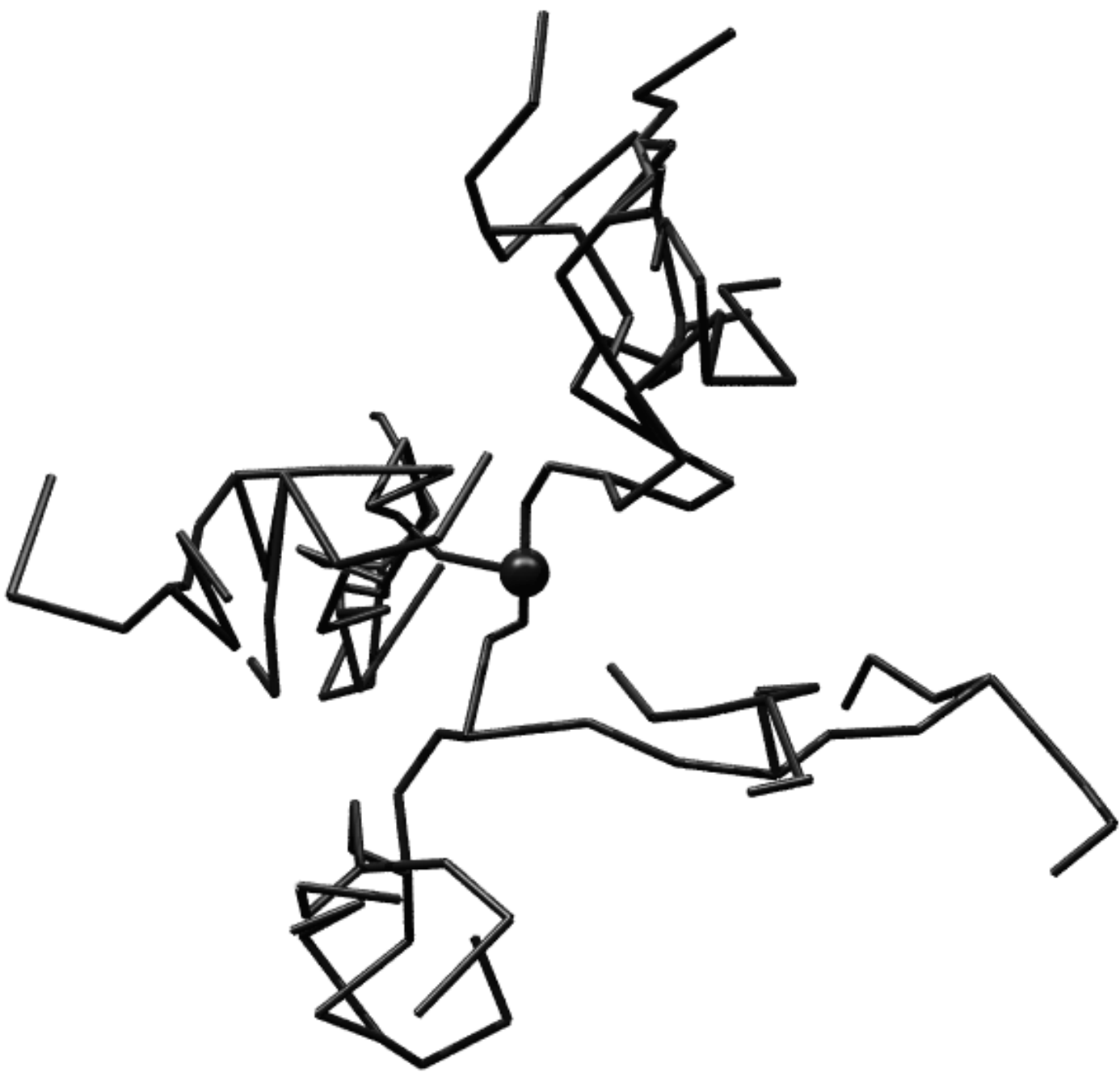
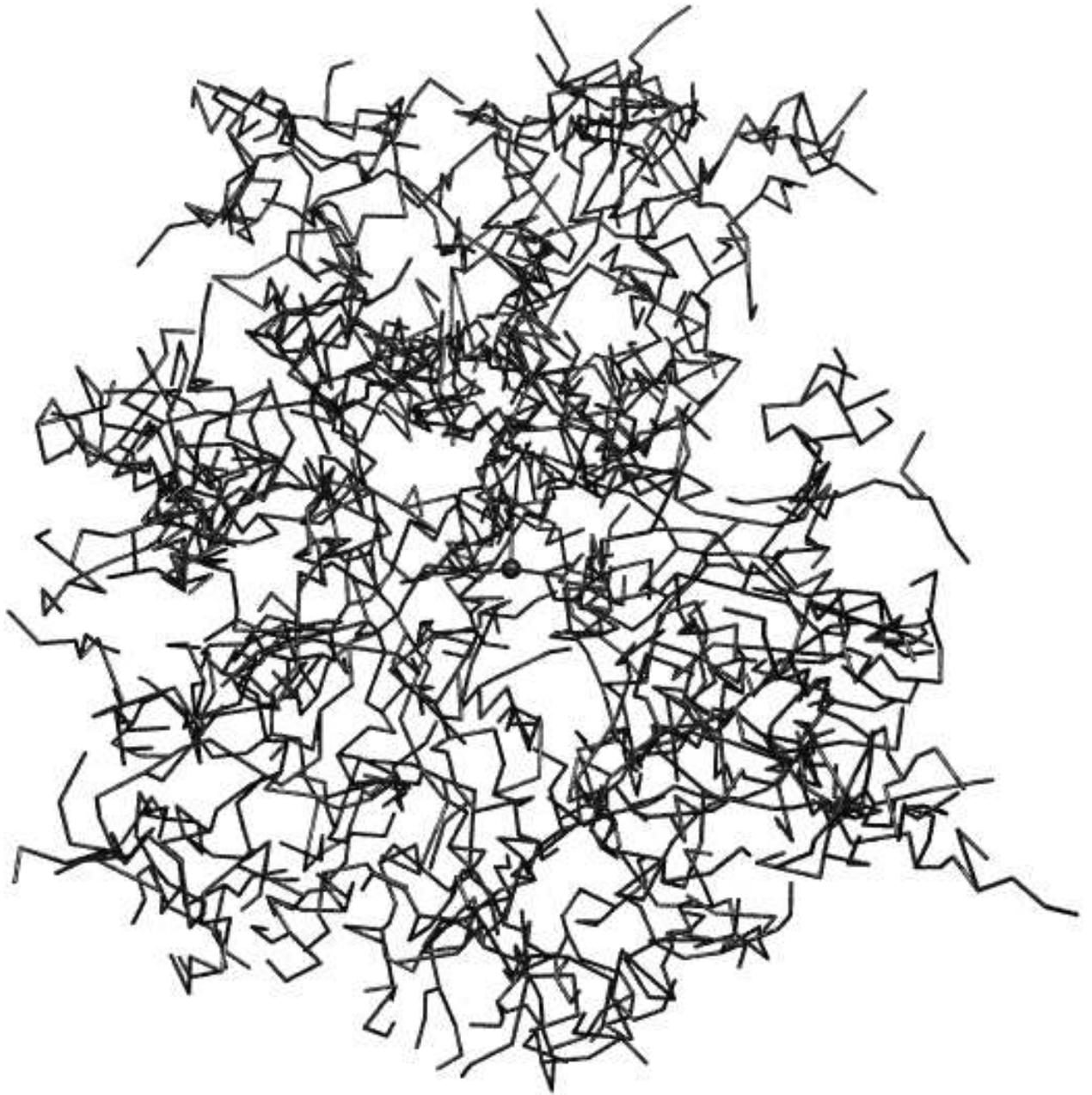
\end{document}